# Optical properties of dense lithium in electride phases by first-principles calculations

Zheng Yu[1], Hua Y. Geng[1]*, Y. Sun[1], and Y. Chen[2]

[1]*National Key Laboratory of Shock Wave and Detonation Physics, Institute of Fluid Physics, CAEP; P.O.Box 919-102 Mianyang, Sichuan P.R.China, 621900*

[2]*Fracture and Reliability Research Institute, School of Engineering, Tohoku University 6-6-01 Aramakiaoba, Aoba-ku, Sendai 980-8579, Japan*

---

* Correspondence and requests for materials should be addressed to H. Y. G. (s102genghy@caep.cn).

# ABSTRACT

The metal-semiconductor-metal transition in dense lithium is considered as an archetype of interplay between interstitial electron localization and delocalization induced by compression, which leads to exotic electride phases. In this work, the dynamic dielectric response and optical properties of the high-pressure electride phases of cI16, oC40 and oC24 in lithium spanning a wide pressure range from 40 to 200 GPa by first-principles calculations are reported. Both interband and intraband contribution to the dielectric function are deliberately treated with the linear response theory. One intraband and two interband plasmons in cI16 at 70 GPa induced by a structural distortion at 2.1, 4.1, and 7.7 eV are discovered, which make the reflectivity of this weak metallic phase abnormally lower than the insulating phase oC40 at the corresponding frequencies. More strikingly, oC24 as a reentrant metallic phase with higher conductivity becomes more transparent than oC40 in infrared and visible light range due to its unique electronic structure around Fermi surface. An intriguing reflectivity anisotropy in both oC40 and oC24 is predicted, with the former being strong enough for experimental detection within the spectrum up to 10 eV. The important role of interstitial localized electrons is highlighted, revealing diversity and rich physics in electrides.

PACS number(s): 78.20.Ci, 78.40.Kc, 78.40.Fy, 61.50.Ks

Keywords: high pressure; metal-insulator transition; lithium; optical property; first-principles

# INTRODUCTION

Pressure alters the state of matter by squeezing electrons into smaller space and modifying the interaction between particles. It is not surprising that metals are usually anticipated to become more "metallic" at high pressures due to expected broadening of the bandwidth and ensued overlapping of electronic orbitals. However, the high-pressure behavior of matter sometimes is counterintuitive and convoluted. Lithium, as the simplest metallic element, is an archetype of how pressure turns a metal into insulator and induces intricate and fascinating phenomena. Since 1960s, the unusual increase in the electric resistivity of lithium at a hydrostatic pressure of 30 GPa as observed in the Bridgman anvil[1,2], and at dynamic 60 and 180 GPa as observed by shock compression method[3,4], has attracted considerable attention, for it indicates that lithium might deviate from the usually expected simple metal behavior. In fact, the bcc structure of lithium at ambient conditions will convert to an fcc structure at 7 GPa under compression, which then transforms to a low symmetry cubic polymorph phase cI16 through an intermediate rhombohedral structure hR1 at about 39 GPa, as discovered by synchrotron X-ray powder diffraction experiments[5].

With pressure increased, the Fermi surface of lithium is substantially distorted by excitation of electrons to $p$ orbitals and making $s$-$p$ hybridization, which then leads to an extended nesting structure, as well as the consequent phonon softening[6,7]. This gives rise to an anomalous melting in lithium compared to other ordinary metals, in which the melting temperature increases to a maximum and then decreases continuously under further compression starting from 520 K at ~10 GPa in the fcc phase[8,9]. A minimum of the melting curve is reached at 50 GPa in the semi-metallic cI16 phase, with a cold melting point as low as 190 K[8], the lowest one among all known materials at this pressure range. The role of quantum ion dynamics in lithium is evident, considering which the accurate theoretical value of melting temperature was obtained as well[10].

Along with this anomaly, the electron-phonon coupling in lithium becomes strong when above 25 GPa even in the fcc phase. This results in one of the highest superconducting temperature (~20 K) of elements[11-15]. The physics behind these peculiar phenomena is that the Coulomb attraction of nucleus is reduced greatly for valence electrons at high pressure, as they suffer a strong repulsion from core electrons due to the overlap with the inner orbitals, as well as the required orbital orthogonality by quantum mechanics. Some valence electrons thus move from the neighborhood of ions to the interstitial positions of crystalline structure, and behave as anions, thus forming high-pressure electrides (HPE)[16-19]. These localized valence electrons are further enhanced by multi-centered chemical bondings that hybridize the valence and conduction bands, and then split them into bonding and anti-bonding states[20]. This exotic behavior was proposed for energy gap opening in dense lithium[17,21-23], and was confirmed by a recent measurement in which a direct visual observation of the reflectance change and a distinct jump in the electric resistivity clearly reveals the existence of an insulating phase of oC40[24]. Under further compression, the multi-centered bonding/anti-bonding states overlap again and lithium returns into a semi-metallic electride phase of oC24 above 110 GPa[22,25,26,27]. Similar anomalous phenomena also present in other light alkali metals and dense hydrogen[28-30]. Especially the *s* to *p* electronic excitation in hydrogen has been predicted to result in a deep melting minimum[31,32] and a mobile solid phase[33]. The interstitial electrons are also shown to be crucial to mix the unmixable lithium and sodium to form an interesting insulating electride alloy[34].

The prominent variation of the atomic and electronic structure in dense lithium induced by pressure should also reflect in physical properties, especially in the vicinity of the boundaries of the metal-semiconductor-metal transition. Dynamic response and optical properties are indispensable for the purpose to acquire a thorough understanding of the high-pressure anomalies in dense lithium[34]. They could provide important information on exotic collective and excitation behavior of valence electrons under an external electric field or interacting with light, which otherwise are not obvious from the bare band structure. These include the zone boundary collective state (as in the bcc phase) and the undamped interband low-energy plasmon that leads to a suppression of the reflectivity (as in the fcc phase) for lithium[35-38]. Such kind of pressure-induced *interband plasmon*, is not unique to lithium[39,40], and cannot be described within the free-electron model[41].

So far, most theoretical calculation and analysis on dense lithium mainly focused on the static crystalline structure and ground-state band structure. The optical properties beyond fcc phase have not been systematically analyzed yet. Considering the drastic change in the crystalline and electronic structure through the metal-semiconductor-metal transition, it is imperative to explore how the optical properties of these HPE phases will evolve under compression. Especially, whether the localized interstitial electrons contribute significantly to the dielectric response? And how it will change at higher pressure beyond 100 GPa? In this work, we comprehensively investigate the dynamic dielectric response and optical properties of lithium at high pressure by first-principles calculations, mainly concentrating on the three intriguing HPE phases of cI16, oC40 and oC24. The intermediate phase oC88 is not considered here for its narrow pressure range of stability[22]. The structural details about these high-pressure phases will be covered in next section. We will then explore the optical properties of lithium from 40 to 200 GPa thoroughly, including the reflectivity, the electron energy-loss spectroscopy (EELS) and the optical conductivity. Both the intraband and interband contributions are carefully treated with the linear response theory. The property in the low energy regime of the spectra, especially within the visible light range, is highlighted for its importance and tight connection with experiments. Ground-state electronic structure and analyses are reported as well, which are prerequisite to interpret the optical spectra. Our investigated pressure range overlaps the transition point of the *2s→2p* excitation. Therefore, the influence of this excitation on the *s-p* hybridization, as well as the impact on the optical properties, are also discussed.

# RESULTS

## A. Weak metallic electride phase cI16

cI16 belonging to the space group $I\bar{4}3d$ is a distorted phase of bcc by separating one pair of lithium ions on the diagonal by a distance of $x$. The atoms locate at the *16c* Wyckoff positions, so that one internal coordinates $(x,x,x)$ is sufficient to represent the crystalline structure uniquely. For example, it takes $(0.05,0.05,0.05)$ at 39.8 GPa[5]. cI16 is a HPE phase and the interstitial electrons locate at the vertices of an equilateral triangle perpendicular to the other diagonal of the bcc structure as shown in supplementary Fig. s3. The electronic structure of cI16 is unique (Fig. 1). It not only exhibits a minimum of density of states (DOS) at the Fermi level, which is a characteristic of semi-metal, but also has almost degenerate bands near the Fermi surface, occurring along $\Gamma - H$, $N - \Gamma$, $\Gamma - Z$ in the reciprocal space. These bands are adjacent to each other when crossing the Fermi level, and lead to low-energy interband transitions, which should be reflected in the low frequency range of the optical spectra.

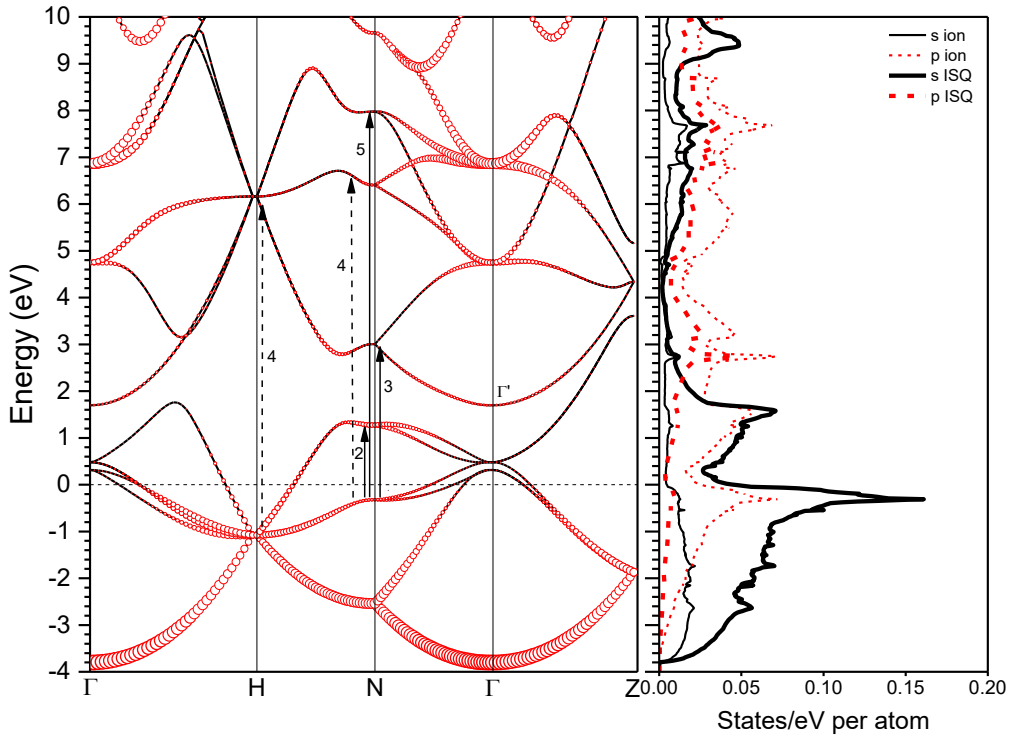

FIG. 1. Calculated electronic band structure and density of states of cI16 at 40 GPa. The circle size in the left panel is proportional to the band *s*-character. The right panel is the corresponding density of states projecting onto an ion and an interstitial quasiatom(ISQ). The arrows (and the numeric numbers) in the left panels indicate the interband transitions that make a major contribution to the respective peaks in the imaginary part of the dielectric function as shown in Fig. 2.

Five distinct peaks can be found in the imaginary part of the interband dielectric function (DF) for cI16 at 40GPa, as shown in Fig. 2(a). The second, third, fourth and fifth peak (at 1.5, 2.8, 7.1 and 7.9 eV, respectively) are consistent with the previous calculations by Alonso et al. [42]. These peaks are mainly associated with the transitions near the N point in the reciprocal space, as the band structure and the high density of the states around this point indicate, which are marked out by arrows in Fig. 1. For example, the transition at N point from the $-0.3$ eV states to the 1.3 eV states matches the position of the second peak at $\sim$1.5 eV very well. For the peak 4, the parallel bands between H and N as the dashed arrows indicate in Fig. 1 also enhance the magnitude of the dielectric response. The discrepancy of the first interband peak with Ref. [42] (possible reasons are discussed in Supplementary Material) is not significant in physics, because the intraband contribution dominates in this frequency range.

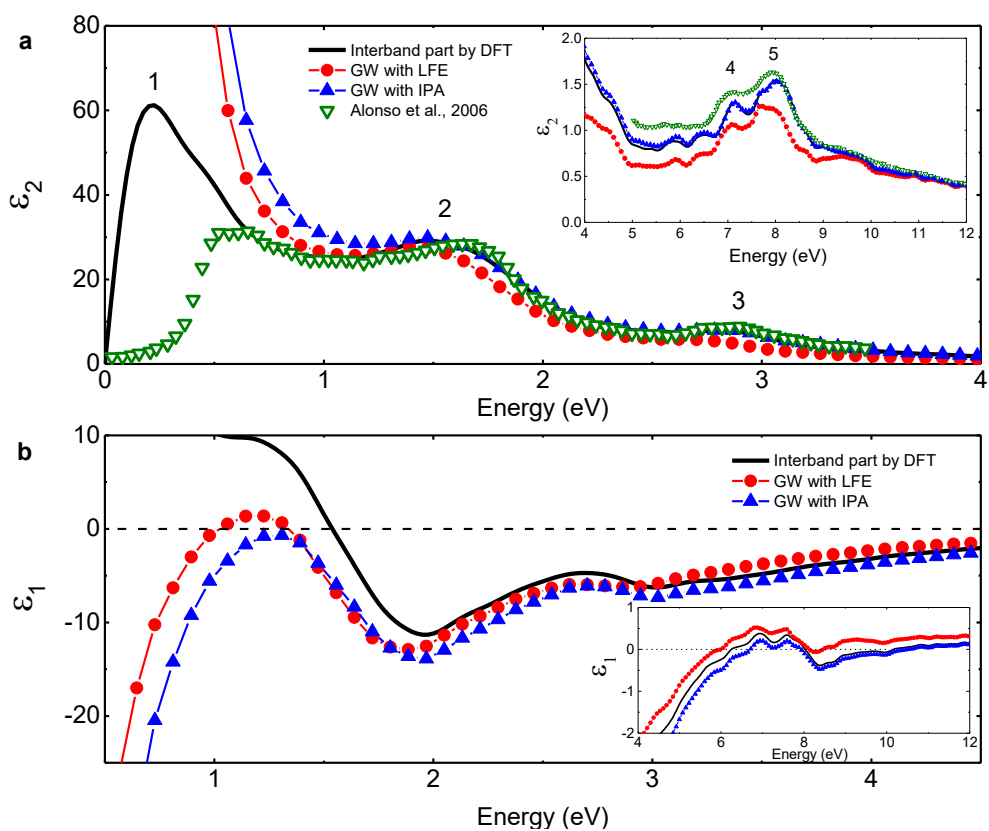

FIG. 2. Dynamic dielectric function (DF) of cI16 calculated with different methods at 40 GPa: (a) the imaginary part and (b) the real part. The solid lines denote the interband part of the DF calculated in the first step by using the DFT method. The lines with filled circles or triangles denote the total DF calculated in the second step by using the GW module, including the LFEs or in the IPA (details in Methods section), respectively. The inset panels show the DF in a wider frequency range of 4-12 eV. The open triangles are the computational results of the interband DF by Alonso et al.[42].

The cI16 phase at 40 GPa is also taken as an example to demonstrate the contribution of intraband transitions and the crystalline local field effect (LFE). As Fig. 2 shows, the characteristics of metal in cI16 phase of lithium are evident when below 1 eV. It has a strong intraband contribution which can be described by the Drude model. However, the intraband contribution is negligible when above 1.5 eV. The LFEs are determined to be small but important in cI16, which move the whole spectra slightly to lower frequency. They even make the value of $\varepsilon_1$ become positive at ~1.2 eV and modify the frequency of the interband plasmon at ~8.2 eV by 2 eV as shown in Fig. 2(b).

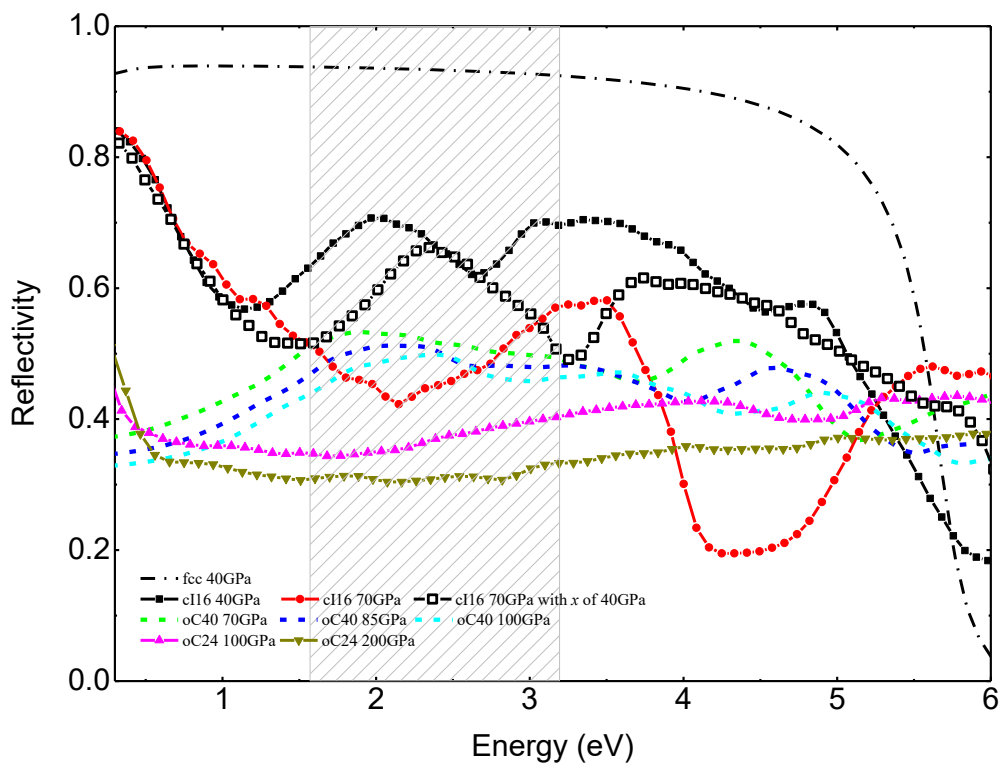

FIG. 3. Variation of the calculated reflectivity of dense lithium across the three HPE phases under increasing pressure as a function of frequency. The shadow region denotes the visible light regime. The line with open squares was obtained by using a structure of cI16 at 70 GPa but fixed the internal atomic coordinate x to that value of 40 GPa. It represents an approximate intermediate stage of cI16 evolving from 40 GPa to 70 GPa. The drastic deviation from the results of the optimized cI16 at 70 GPa highlights the importance of the local atomic structure to the optical properties.

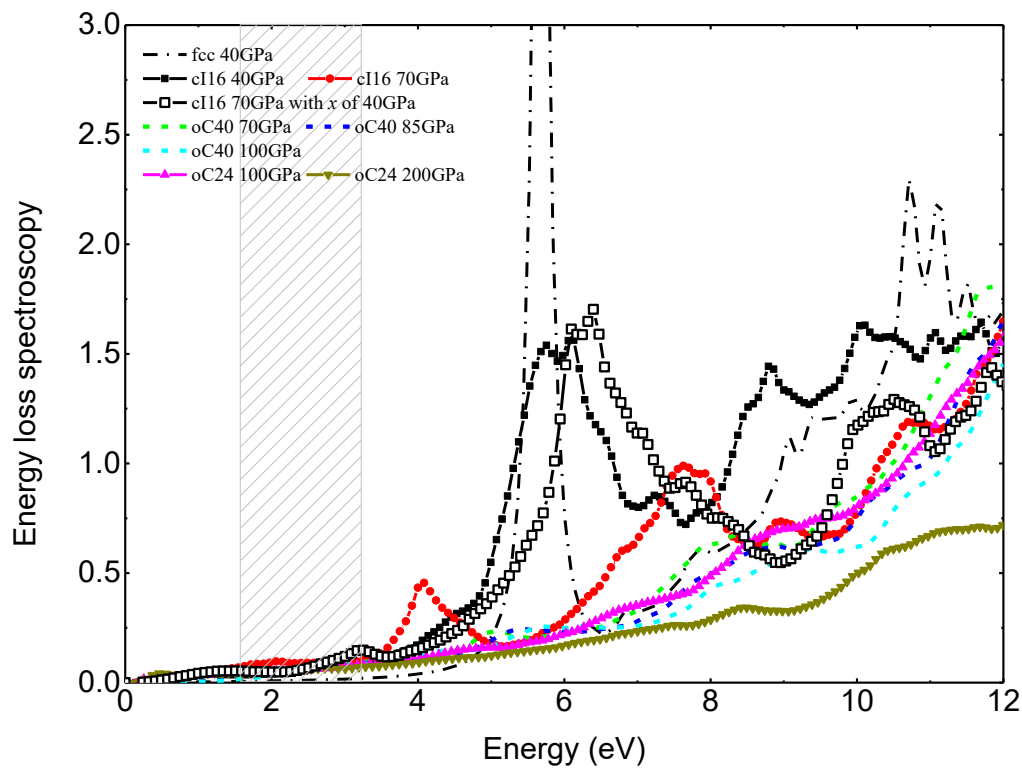

FIG. 4. Variation of the calculated electron energy-loss spectroscopy (EELS) of dense lithium across three HPE phases under increasing pressure as a function of frequency. Note the splitting of the main EELS peak in cI16 phase into two smaller ones when pressure was increased from 40 GPa to 70 GPa, in view of which the importance of the local structure optimization is evident.

The reflectivity and the EELS of cI16 and fcc phase at 40 GPa were calculated and shown in Figs. 3 and 4. Our reflectivity data of the fcc phase agree with the calculations by Silkin et al.[35], where the reflectivity has a value above 0.9 up to 5.8 eV and is then strongly suppressed by the interband plasmon at 6.0 eV. The reflectivity of cI16 by our calculations in the infrared region is about 0.3 smaller than fcc, namely, reduced by 40%, due to the LFE-induced small value in $\varepsilon_1$ at 1.2 eV. This is in qualitative agreement with the previous experiment that the reflectivity in visible near infrared range exhibits a jump when transition from fcc to cI16 phase[43]. The subtle details of the computed dynamic spectrum at 40 GPa, however, is a little bit different from the experiment at 36 GPa, as detailed in supplementary Fig. s4. Optical properties sensitively depend on atomic structure, as revealed in Fig. 1. There is a complex phase transition sequence of fcc → hR1 → cI16 in lithium at around 36~40 GPa. The inevitable pressure gradient across the sample chamber in the DAC experiment also could lead to multi-phase coexistence, as well as polycrystallization. All these might contribute to the numerical deviations.

The electronic structure of cI16 changes a lot under further compression, especially at Γ point (see supplementary Fig. s5). Increasing the pressure to 70 GPa moves up the states marked as Γ′ in Fig. 1 dramatically and forms a new triply degenerate state at ~4.5 eV, resulting in a sharp maximum edge of the density of states at about 4 eV. The lifting and broadening of conduction bands, especially those around N point, shifts the peak 1∼5 in the imaginary DF to higher frequency. Their respective amplitudes are accordingly reduced, except for the peak 4 and 5. The amplitude increase of the peak 4 and 5 might be due to the corresponding parallel states along the H-N direction. Noticeably, a new peak is split out from the first peak at 0.7 eV at 70 GPa. This corresponds to the transitions at the midway between the N and Γ points. The overall trend in the imaginary part of DF is that the whole spectrum gradually becomes flat. However, it always obeys the optical f-sum rule[44]

$$\int_0^\infty \varepsilon^{(2)}(\omega)\omega d\omega = \frac{2\pi^2 e^2 n}{m} = \frac{\pi}{2}\overline{\omega}_{Drude}^2(n) \,. \tag{1}$$

In the visible light range (1.6-3.2 eV), the reflectivity of cI16 is strongly modified by pressure so that it is almost 0.3 smaller at 70 GPa than that at 40 GPa (or 43% reduction). This low-energy suppression originates in the shift of the intraband plasmon from 1.2 eV to 2.1 eV by compression (see Fig. 5(b)). At 40 GPa, the strong plasmon peak at about 5.8 eV in EELS for the fcc phase is broadened when transition into the cI16 phase. It then splits into two smaller peaks at 70 GPa with one locates at 4.1 eV, and another one at 7.7 eV. This splitting is due to the moving up of the ε₁ peak between 2 and 3 eV at 40 GPa (Fig. 5(b)) to 4.8 eV at 70 GPa, and crossing the zero line at 4.1 and 7.7 eV, respectively. As a result, two new interband plasmons emerge and substantially suppress the reflectivity of cI16 at the corresponding frequency (the strong suppression at ∼4.1 eV in cI16 at 70 GPa is evident in Fig. 3. The similar suppression at ∼7.7 eV is not shown here).

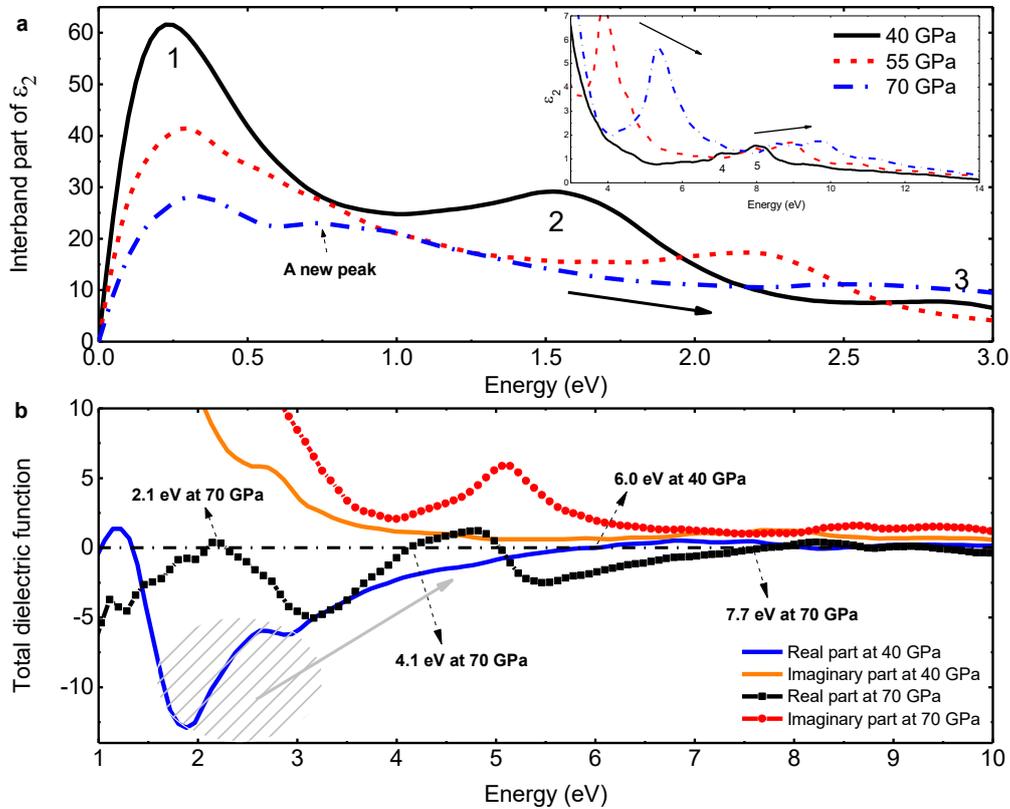

FIG. 5. (a) Variations in the interband part of the imaginary dielectric function of cI16 induced by pressure. The solid

arrows show the trend of peaks shifting with increasing pressure. A new peak at 0.7 eV emerges at 70 GPa as the dashed arrow indicates. (b) The real and imaginary parts of the total dielectric function in cI16 at 70 GPa, compared to those at 40 GPa. The plasmon frequencies are also labeled. The shadow region highlights the 2-3 eV part of the real DF at 40 GPa, which shifts in the direction as the long arrow shows and induces two new plasmons at 4.1 and 7.7 eV under further compression.

We hypothesized the anomalous variation in the optical properties induced by compression to 70 GPa, especially the splitting of the plasmon at 6.0 eV and the emergence of the weak intraband plasmon at ~2.1 eV, might arise from the local distortion characterized by the internal coordinate $x$, rather than the hydrostatic compression, according to two facts: (1) this $x$ value in cI16 phase was analyzed and found significant to optical spectra by Alonso et al., but only in the range of 0.01-0.05[42]; (2) in our calculation, the value of $x$ actually increases further under compression (e.g., $x$=0.048 at 40 GPa, 0.060 at 55 GPa, 0.076 at 70 GPa). To validate this assumption, we separate the structure change from 40 GPa to 70 GPa into two steps. In the first one, the lattice vectors of the structure are released to 70 GPa but with the internal coordinate $x$ being fixed at the value of 40 GPa, i.e., $x$=0.048. In the second step, the internal coordinate is also fully optimized. The reflectivity of this intermediate phase shows an overall agreement with that of 40 GPa, with just a slight shift of the spectrum towards higher frequency and reduction in the magnitude, as shown in Fig. 3. This similarity is also observed in EELS. It unequivocally demonstrates that the major contribution to the anomalies in optical property variation comes from local structure distortion. This structural deformation should partially associate with the increase of the $p$-contribution to the electronic states around ions, accompanying which, the $p\pi$ bonding in cI16 becomes more saturated, allowing more valence electrons to localize at interstitial space. The enhanced localization of interstitial electrons by compression gradually requires more void space for accommodation. Thus, the structure of cI16 is distorted with the increment of the internal coordinate $x$. A larger value of $x$ indicates stronger deviation from the initially high symmetric phase. This process is more drastic in cI16 than in the other two electride phases, due to its evident localization change under compression. For example, the volume of interstitial space with electron localization function (ELF) greater than 0.9 increases five times from 40 to 70 GPa in cI16, compared to a mild increase of 18% in oC40 when compressed from 70 to 100 GPa (see Supporting Information).

## B. Insulating electride phase oC40

The phase transition sequence in lithium at ~60 GPa is cI16 → oC88, which further changes into oC40 at ~68 GPa at a temperature of about 200 K, as reported by experiment[8]. The same transition sequence was also predicted by *ab initio* calculations after including the zero-point energy[22]. The orthorhombic phase oC40 belongs to the space group of *Aba2*. It is the only phase that has an energy gap in all intrinsic phases of dense lithium, as shown in Fig. 6. Because of its insulating characteristic, there is no intraband contribution, and the band summation method equation (8) can be directly applied, which corresponds to the interband part of optical properties and is easy to calculate. However, a careful check of the influence of the exchange-correlation functional and the crystalline local field is required. Our calculated results as shown in Fig. 7 obviously show that the band summation method is a good approximation and the LFEs are negligible, especially for the energy range larger than 2.5 eV, where the DFT results overlap with those given by GW module.

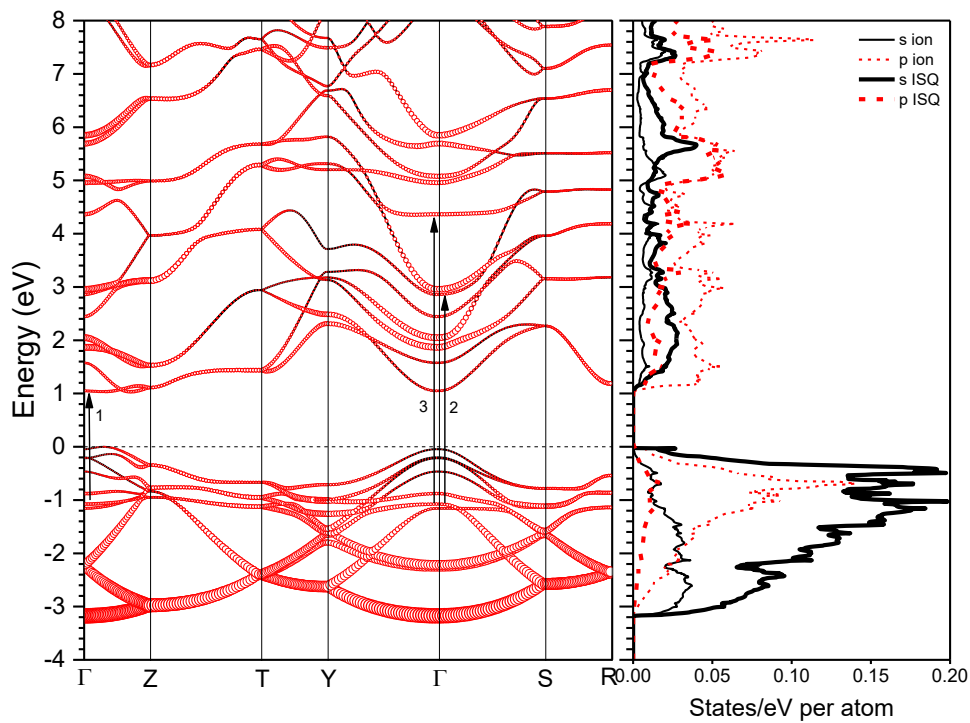

FIG. 6. Calculated electronic band structure and projected density of states of oC40 at 85 GPa. The circle size is proportional to the band *s*-character in the left panel. The arrows in the band structure denote the interband transitions that make a major contribution to the respective peaks of the imaginary DF as shown in Fig. 7.

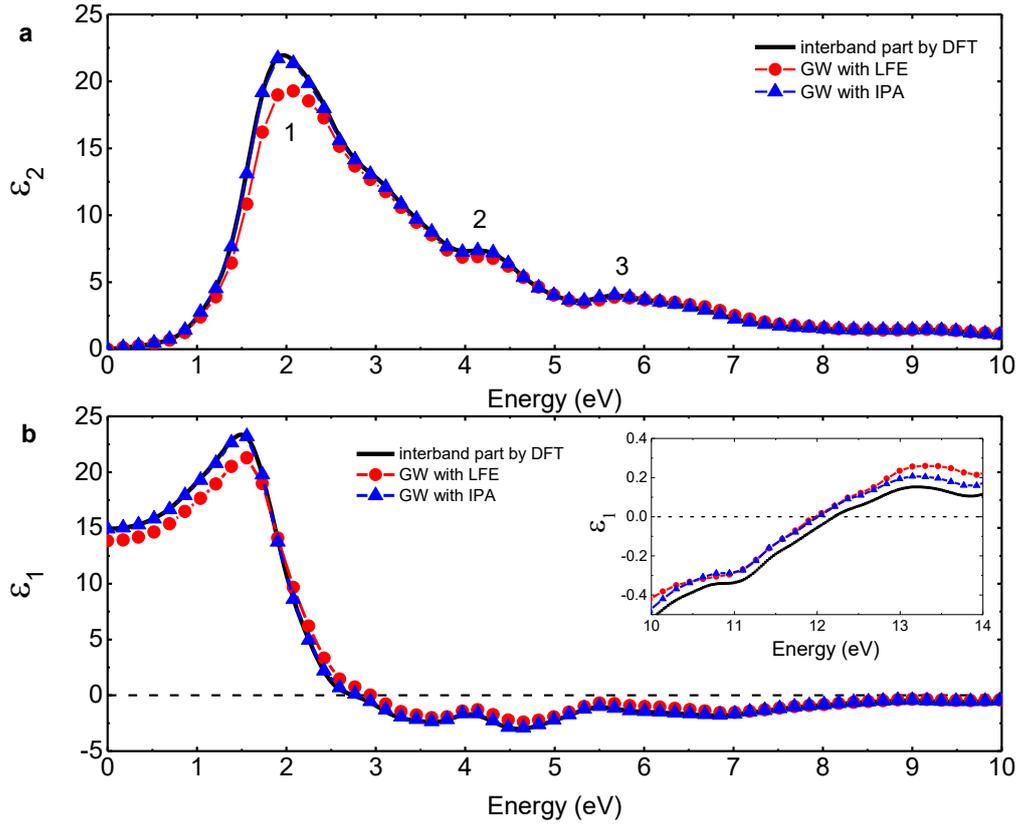

FIG. 7. Dynamic dielectric function of the oC40 phase calculated with different methods at 85 GPa. The reported DF is spherically averaged by $\varepsilon_a^{(2)} = (\varepsilon_{xx}^{(2)} + \varepsilon_{yy}^{(2)} + \varepsilon_{zz}^{(2)})/3$. The solid lines denote the interband part of the DF calculated in the first step by using the DFT method. The lines with filled circles or triangles denote the DF calculated in the second step by using the GW module, including the LFEs or in the IPA (detailed in Methods section), respectively.

From first-principles band structure calculations, we find an indirect band gap in oC40 about 1.03 eV (1.60 eV in the GW calculation) at 85 GPa. As a consequence, both the imaginary part of DF and the absorption spectrum show a feature starting from 1 eV. As shown in Fig. 7, three evident peaks appear in the imaginary DF. The main peak at 2.1 eV exactly relates to the transitions at Γ point from the states of −1 eV to 1 eV. They correspond to the first peak above and below the Fermi level in the electronic DOS, respectively. The second peak at 4.1 eV is due to the transitions at Γ point from the states of −1 eV to 3 eV, and the third one at 5.6 eV originates in the transitions within a wider momentum range along the Y − Γ − S direction and centered on Γ point in the Brillouin zone (BZ) from the states of -1 eV to 4.6 eV. The real part of the DF crosses the zero line at 2.9 eV and 12 eV, respectively.

The symmetry of oC40 becomes lower compared to cI16, both in the crystalline structure and electronic distribution. The conventional cell of oC40 can be divided into two centrosymmetric parts along *b* axis, with each part including three different layers. Between these two parts is the largest interstitial space in oC40, where localized valence electrons locate, noted as M1 and M3. The other interstitial localization position is in the middle layer of each part, noted as M2 (see supplementary

Fig. s7). This inhomogeneous characteristic gives rise to an anisotropy of optical properties for oC40. In our calculation, although the imaginary parts of the DF in oC40 share a main peak at ∼2 eV, the amplitudes of this peak with polarization along *a* and *c* axis are almost twice as that along *b* axis, as shown in Fig. 8(a). This is directly due to an inhomogeneous distribution of *x*, *y*, and *z* components (65% for $p_y$ at 1eV) in the *2p* orbital of the conduction state. The second peak with light polarization along *c* axis has a slight shift compared to that along *b* axis, and they coincide with each other at the third peak. The reflectivity anisotropy is more evident as shown in Fig. 8(b). A difference of 0.1 is observed in the whole infrared and visible light regime (specific numbers are shown in Table I). For example, at 2.5 eV the reflectivity along *b* axis is 0.17 smaller than that along *a* axis. In ultraviolet regime, the reflectivity even increases 62% when changing the light polarized direction from *b* to *c* axis.

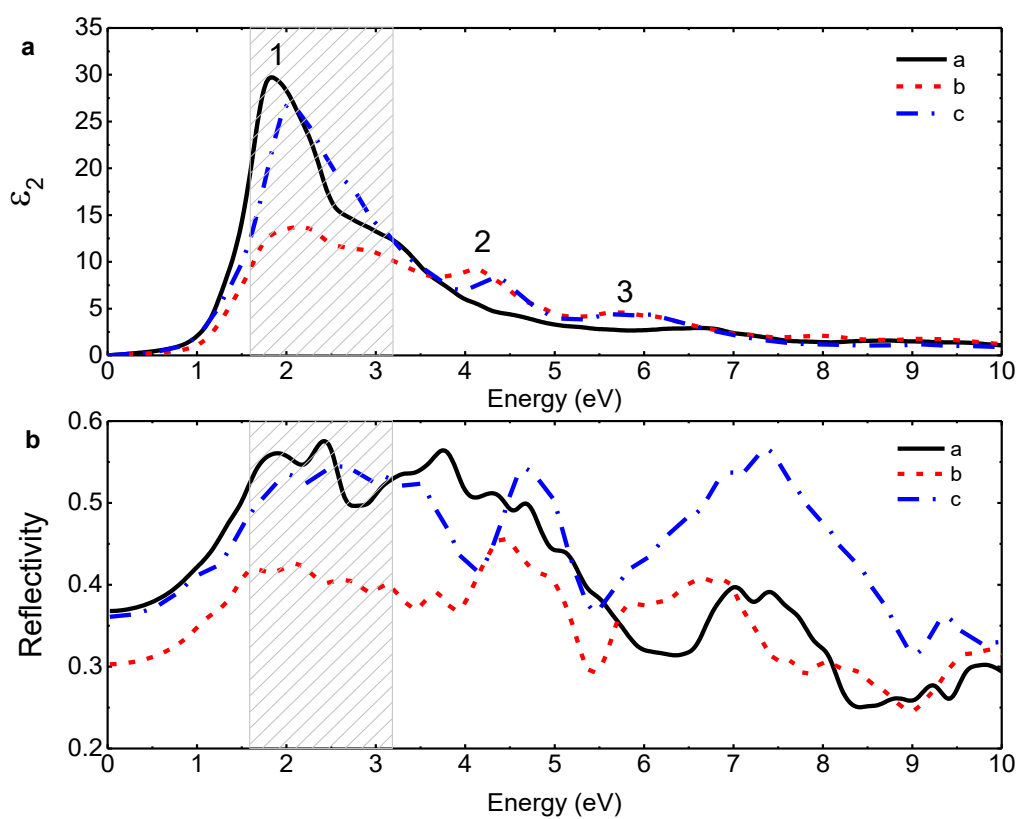

FIG. 8. Calculated anisotropy of optical properties in oC40 at 85 GPa: (a) the imaginary part of the DF and (b) the reflectivity with different light polarizations along *a*, *b* and *c* axis, respectively. The peaks in Fig. 7 are marked in (a) as well. The shadow region denotes the visible light range. There is a maximal percentage difference of 29% at 2.5 eV and 62% at 7.4 eV based on the reflectivity along *a* axis.

Under compression, the energy of the *2s* orbital in lithium increases more quickly than that of the *2p* orbital. This can be easily understood with quantum mechanics that the wavefunction of the *2p* orbital has one less nodes than the *2s* orbital in radial direction. For the insulating oC40 phase, the bottom of the conduction bands essentially has more *s* component than the top of the valence bands. As a result of this, the valence bands are squeezed to a narrower energy range (3.5→2.8 eV) and the calculated band gap increases at a rate of ~0.013 eV/GPa when pressure rises from 70 to 100 GPa. The features in optical spectra correspondingly move to higher frequency and the reflectivity is slightly reduced under continuing compression. Unlike the local structural distortion observed in cI16, all pressure-induced variations in oC40 seem mild.

TABLE I. The reflectivity of high-pressure phases in lithium from 40 to 200 GPa at a frequency of 2, 3 and 4 eV, respectively. The optical properties of oC40 at 85 GPa are provided with different light polarizations.

| Phase | Pressure (GPa) | 2eV | 3eV | 4eV |
|---|---|---|---|---|
| fcc | 40 | 0.94 | 0.92 | 0.90 |
| cI16 | 40 | 0.71 | 0.70 | 0.65 |
|  | 70 | 0.45 | 0.54 | 0.30 |
|  | 70 ($x$=0.048) | 0.60 | 0.56 | 0.61 |
| oC40 | 70 | 0.53 | 0.50 | 0.49 |
|  | 85 (along *a* axis) | 0.56 | 0.51 | 0.52 |
|  | 85 (along *b* axis) | 0.42 | 0.39 | 0.38 |
|  | 85 (along *c* axis) | 0.53 | 0.53 | 0.42 |
|  | 100 | 0.49 | 0.46 | 0.44 |
| oC24 | 100 | 0.35 | 0.40 | 0.43 |
|  | 200 | 0.31 | 0.32 | 0.36 |

## C. Reentrant weak metallic electride phase oC24

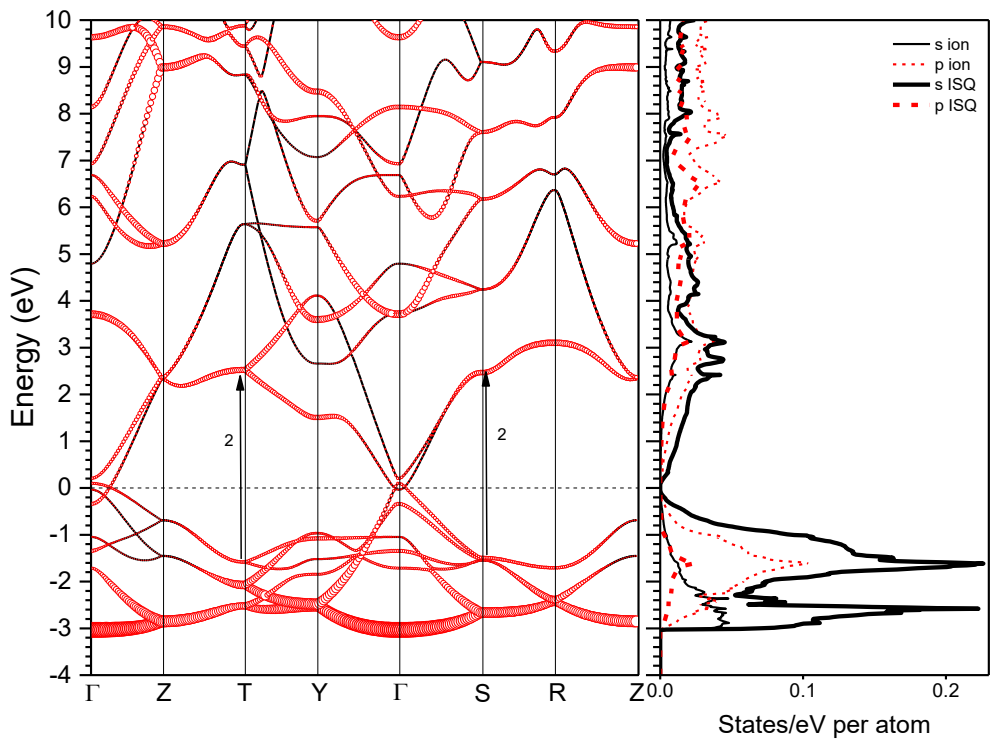

FIG. 9. Calculated electronic band structure and projected density of states of oC24 at 100 GPa. The vertical arrows denote possible interband transitions that make a major contribution to the second peak in the imaginary DF as shown in Fig. 10(a). Others are the same as Fig. 6.

Further compression leads to band overlap again, and lithium transforms into a reentrant metallic electride phase oC24. The atomic structure of this phase, belonging to the space group *Cmca*, was first predicted by Pickard using *ab initio* method[18]. The metallic feature of this phase was experimentally confirmed by a rapid decline in electric resistivity at ∼110 GPa[25]. By comparison to the semi-metal phase cI16 at low pressure, this high-pressure phase has an almost vanished electronic DOS at the Fermi level ($E_F$). As shown in Fig. 9, a Dirac-cone-like structure that represents a nearly linear dispersion relationship appears around the Γ point in the reciprocal space. This feature makes oC24 lithium an extremely weak metal, and there is almost no intraband contribution. The calculated imaginary part of DF at 100 GPa reveals two main peaks at 0.2 eV and 3.9 eV, respectively (see Fig. 10(a)). The first one corresponds to the interband transitions within the cone around the Γ point, and the second one mainly originates from the transitions of −1.5 → 2.5 eV at S and T points in the BZ, as well as the −0.4 → 3.65 eV transition at the Γ point. The shoulder peak at 5.2 eV is mainly due to the −0.4 → 4.8 eV transition at the Γ point. The real part of DF crosses the zero line at 4.2 and 13.5 eV, respectively. However, because of the large magnitude of the second main peak around 4 eV, only at 13.5 eV there forms a strong interband plasmon. The local field effects are investigated and found noticeable in oC24. The position of the second main peak in the imaginary DF and the locations of zero in the real part of DF are moved to lower energy by ∼1 eV due to the LFEs. The dielectric response in this phase has perceptible anisotropy, which however is weaker than oC40, especially in the visible light range (detailed in supplementary Fig. s10).

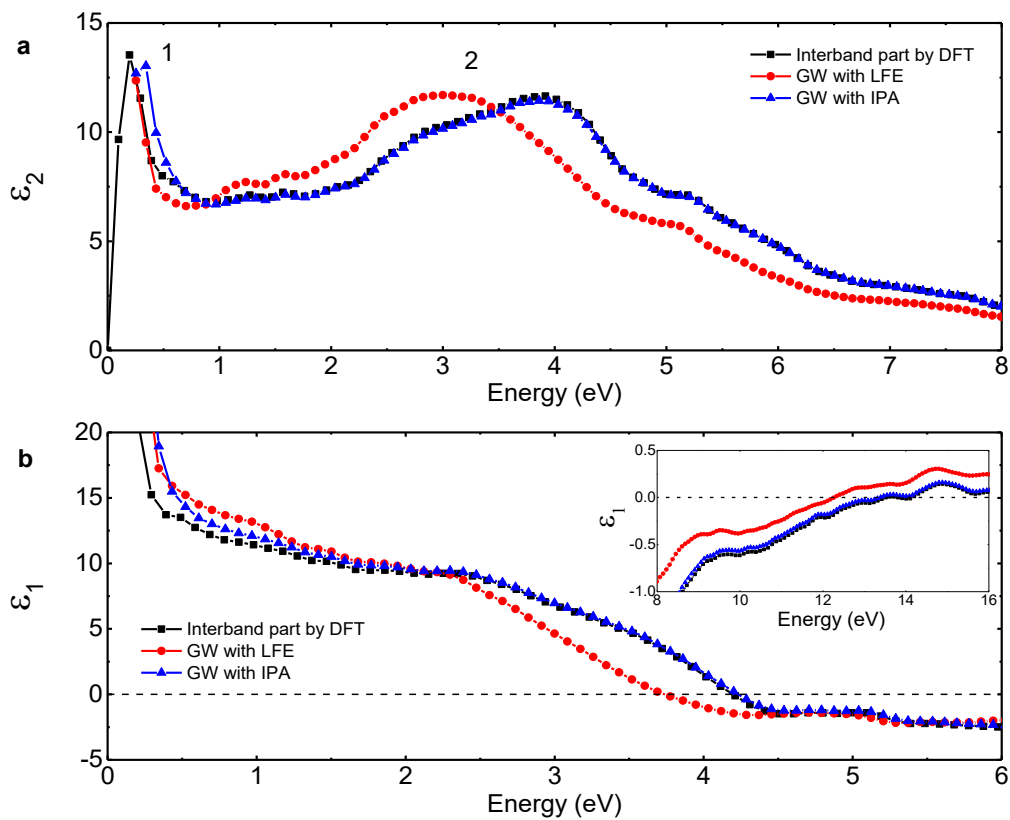

FIG. 10. Spherically averaged dynamic dielectric function of oC24 calculated with different methods at 100 GPa. Other settings are the same as Fig. 7.

In view of the low frequency part of the calculated $\varepsilon_1$ (see Fig. 10(b)), oC24 is more like an insulator that allows low-frequency electromagnetic wave to propagate, rather than an electronic shielding metal like fcc or cI16 phase. Furthermore, the large positive value of $\varepsilon_1$ in oC24 extends a wide frequency range up to 4 eV, compared to the drastic drop of $\varepsilon_1$ at 2 eV in oC40, which means that visible light is reflected less and can transmit more easily in oC24 than in oC40. This explains why oC24 has the lowest reflectivity and absorption in the visible light range among these three electride phases, as shown in Fig. 3 and Table s1. In low frequency range, the imaginary part of DF for oC24 has large positive values like other metals, which makes its optical conductivity much larger than oC40. It is interesting to point out that under further compression, the electronic band structure around the Fermi level, especially the cone feature, keeps almost unchanged even up to a pressure of 200 GPa. Other part of the band structure, however, changes gradually with increasing hydrostatic pressure, just like the situation in oC40 (see supplementary Fig. s11). Consequently, the reflectivity and the EELS decrease further, and dense lithium becomes more transparent.

# DISCUSSION

On the whole, the reflectivity of lithium in the visible light range decreases as a function of pressure. It changes slowly within one stable phase and rapidly, even a jump at the transition pressure. Some specific values of reflectivity at given frequencies are provided in Table I. Compared to oC40 and oC24, the reflectivity of cI16 is modified strikingly by pressure. Its reflectivity at 2 eV changes 40% when the pressure is increased by 30 GPa, whereas there are only 8% change for oC40, and less than 11% for oC24 with the same pressure increment. This difference becomes more drastic at 4 eV. Especially, the calculated reflectivity at 3 eV deceases when transition from cI16 to oC40 phase at 70 GPa, in agreement with the experimental observation[24]. The calculated reflectivity also strongly depends on the frequency of incident wave, and this dependence greatly differs from each other in these three electride phases, mainly due to their different plasmon frequencies. As demonstrated by EELS in Fig. 4, the main plasmon frequencies in both oC40 and oC24 are higher than 12 eV, whereas the cI16 phase has pronounced plasmons in low frequency range. For example, it has one intraband plasmon at 2.1 eV and two interband plasmons at 4.1 and 7.7 eV at 70 GPa, respectively. According to equation (2), the reflectivity is closely associated with the inverse of the absolute value of DF, and this value has a relatively evident maximum in the vicinity of the plasmon frequency, thus leading to a strong suppression of the reflectivity. This mechanism clearly explains why the reflectivity of oC40, oC24, and fcc changes only by 0.12, 0.06, and 0.04 when the frequency increases from 1 to 4 eV, whereas it changes remarkably by 0.38 in cI16 at 70 GPa.

The static optical conductivity we calculated as a function of pressure is qualitatively in good agreement with the experimental measurements using DAC[24,25]. For oC24, the calculated static conductivity is lower than cI16, but much higher than oC40 (see Fig. 11), reflecting the strength of their respective metallic characteristics. The jumps at the phase transition boundaries are evident, especially the striking one when transition from cI16 to oC40 phase. The deviation in the numerical value is understandable, such as for the cI16 phase at 40 GPa, since no scattering due to phonons and impurities had been considered in our calculation, which gradually becomes important at low temperature. The possibility of multi-phase coexistence in the vicinity of the phase boundaries could also contribute to the difference. In addition, the high-frequency alternating current (AC) conductivity varies dramatically with frequency so that the metal phases, such as the fcc phase, have a very low conductivity above 1.5 eV, whereas the conductivity of oC40 gradually increases for frequencies above 1 eV and becomes the largest one among the fcc, cI16, and oC24 phases at 2 eV (supplementary Fig. s6).

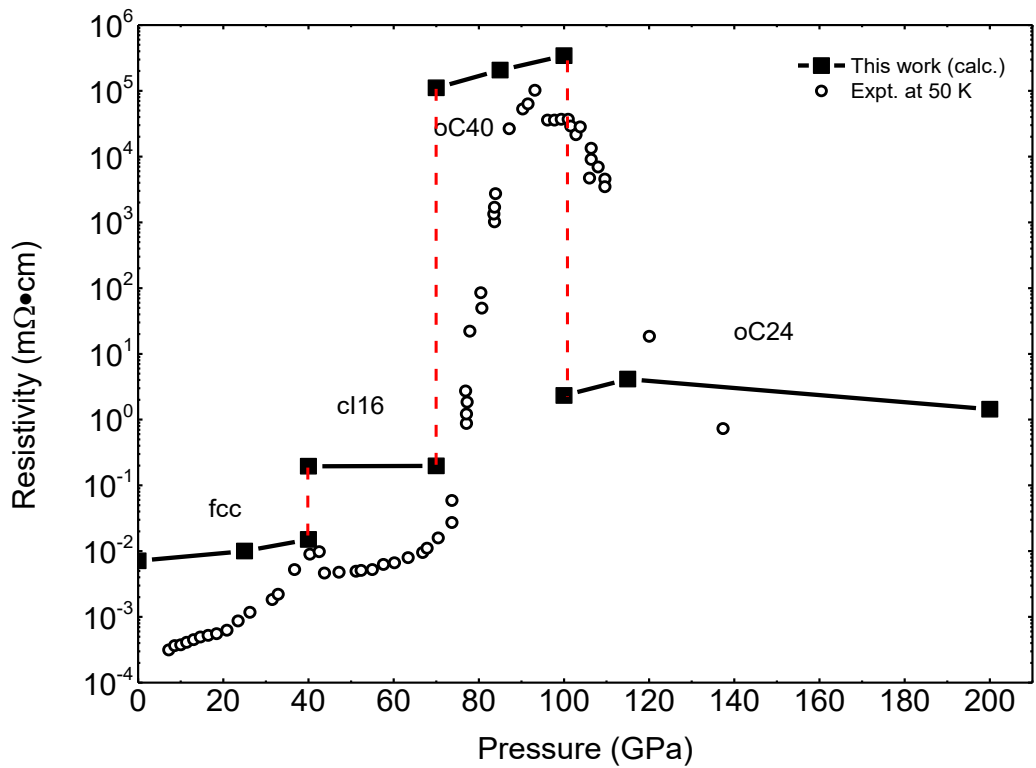

FIG. 11. Comparison of the calculated static electric resistivity as a function of pressure with the experimental data at 50 K[25].

Except the evident variations of the electronic and optical properties at the phase boundaries, which are described by the $s \rightarrow p$ excitation and *s-p* hybridization theory, the origin of compression-induced changes within one stable phase can be accounted by: (i) the regular effect of hydrostatic pressure, as shown in oC40 and oC24, as well as the intermediate step of cI16 discussed in Results section; (ii) the irregular effect that causes extra structural modification, such as the distortion in cI16 at 70 GPa. However, this irregular effect of compression of (ii) only emerges in some particular phases, according to above analysis. We infer that its working condition includes both the low symmetry crystalline structure and the metallic electronic structure with energy bands crossing the Fermi level. The first one is easy to understand since less constraint by spatial symmetry makes stronger modification on crystalline structure possible. Therefore, this effect does not show up in the bcc and fcc phase. The second one is closely associated with the electronic structure around the Fermi surface, because variation of this part impacts essentially on the electron localization in electrides and further modifies all physical properties, especially the optical properties in low frequency range. oC40 and oC24 phase do not have many such bands crossing the Fermi level, so that the role of compression in them belongs to the case (i).

Furthermore, the optical electronic transitions occur only when the angular momenta satisfy the optical selection rule $\Delta l = \pm 1$. Figs. 1, 6, and 9 show the electronic DOS projected onto the lithium ions and interstitial quasiatoms (ISQ) in cI16, oC40, and oC24 phases, respectively. The valence electrons in these electride phases mainly occupy the *s* orbitals centered at ISQs rather than the *p* orbitals surrounding the ions. Consequently, the valence electrons from ISQs, when interacting with light, are inevitably excited to *p* orbitals that are usually surrounding the lithium ions. Because all electride phases are stabilized by the interstitial electrons, as well as the ensuing multicenter-bonding, this photon-induced electron transfer from ISQs to ions implies there might have a strong photon-phonon coupling, leading to polaritons or light-induced phase transformations. Our rudimentary calculation on oC40 with such photon-excited electronic configuration indeed suggests that dense lithium could be a prototype of elements with such kind of photon-induced transition, and should be the focus of future investigations.

To summarize, the dynamic dielectric response and optical properties of the three high-pressure electride phases of cI16, oC40, and oC24 in lithium were thoroughly and comprehensively investigated across a wide pressure range from 40 to 200 GPa by first-principles calculations. The reflectivity of dense lithium was predicted to decrease with increasing pressure in the visible and infrared range, changing from highly reflective metallic phases to semi-transparent electride phases. Two previously unknown interband plasmons were discovered in cI16 at 70 GPa. This explains the abnormal variation in the reflectivity, and reveals a strong dependence of the optical property on the subtle details of the crystalline structure. The atomic structure was found gradually distorted with the increasing localization of interstitial electrons under compression, even within the same cI16 phase. In contrast to the drastic jump of physical properties at the phase boundaries, this observed gradual evolution unveils how external pressure continuously modifies the microscopic structure, especially the irregular effect of compression, via localized interstitial electrons, and further impacts on the dielectric response. These interesting variations, however, are not obvious if viewed only from the band structure. The system accumulates these variations and finally reaches the condition of phase transitions. We also predicted the possible response of electrides when interacting with the light field, and discovered diverse and rich features in their optical properties. Two important characteristics are the strong anisotropy in oC40 phase and the possible photon-induced transformation. Even though both cI16 and oC24 are semi-metallic electrides, their optical properties, as well as the role of compression in their evolution, are very different, which is quite unexpected. Especially, oC24 becomes a semi-transparent metal, with a reflectivity and absorption coefficient even lower than the insulating oC40 phase, determined by its unique electronic structure. This work uncovers intriguing anomalous optical properties in the electride phases of dense lithium, and the underlying physics was elaborated. The results and findings could be the foundation for future experimental exploration of high-pressure electrides.

# METHODS

## A. General aspects of dielectric function calculation

The dielectric function (DF) $\varepsilon(\omega, \mathbf{q})$ is an inherent physical quantity that determines matter's response to an external electric field. It strongly depends on the frequency $\omega$ and the wave vector $\mathbf{q}$ of the incident electromagnetic wave. For $\lim_{\omega \to 0} \varepsilon(\omega, \mathbf{q})$, it describes the electrostatic screening of interactions between electron-electron, electron-lattice and electron-impurity[45]. In this work, the dynamic dielectric function at the long-wavelength limit of $\mathbf{q} \to 0$ is studied, which corresponds to a slowly varying macroscopic electric field, where the longitudinal and transversal components of DF become equal[46]. The response to this electric field reveals a collective excitation of the Fermi sea, or the bulk plasmon behavior.

The important optical properties, such as the reflectivity $R(\omega)$, the EELS $L(\omega)$, the refractive index $n(\omega)$, and the absorption coefficient $I(\omega)$, all can be obtained from the dynamical dielectric response function $\varepsilon(\omega) = \varepsilon_1(\omega) + i\varepsilon_2(\omega)$. The explicit expressions of them are given as[47]

$$R(\omega) = \left|\frac{\sqrt{\varepsilon(\omega)} - 1}{\sqrt{\varepsilon(\omega)} + 1}\right|^2, \qquad (2)$$

$$L(\omega) = \frac{\varepsilon_2(\omega)}{\varepsilon_1(\omega)^2 + \varepsilon_2(\omega)^2}, \qquad (3)$$

$$n(\omega) = \frac{1}{\sqrt{2}}\left[\sqrt{\varepsilon_1(\omega)^2 + \varepsilon_2(\omega)^2} + \varepsilon_1(\omega)\right]^{1/2}, \qquad (4)$$

$$I(\omega) = \sqrt{2}\omega\left[\sqrt{\varepsilon_1(\omega)^2 + \varepsilon_2(\omega)^2} - \varepsilon_1(\omega)\right]^{1/2}. \qquad (5)$$

Here the EELS describes inelastic scattering of the incident electrons with narrow initial kinetic energy, including the inelastic interaction of inter- and intra-band transitions and plasmon excitations. The dynamic optical conductivity $\sigma(\omega) = \sigma_1(\omega) + i\sigma_2(\omega)$ can be calculated by using the dielectric function as[48]

$$\sigma_1(\omega) = \varepsilon_0 \omega \varepsilon_2(\omega), \qquad (6)$$

$$\sigma_2(\omega) = \varepsilon_0 \omega (1 - \varepsilon_1(\omega)). \qquad (7)$$

The most fundamental work for optical property evaluation thereby is to calculate the dielectric function. This is divided into two steps, with different approximations being adopted in each step. In the first step, we employ a method of summing all possible transitions over the whole conduction band after the electronic ground state has been obtained. The imaginary part of the dielectric function is then determined by following equation[49]

$$\varepsilon^{(2)}_{\alpha\beta}(\omega) = \frac{4\pi^2 e^2}{V} \lim_{q \to 0} \frac{1}{q^2} \sum_{c,v,\mathbf{k}} 2w_{\mathbf{k}} \delta(\varepsilon_{c\mathbf{k}} - \varepsilon_{v\mathbf{k}} - \omega) \times \langle u_{c\mathbf{k}+e_\alpha q} | u_{v\mathbf{k}} \rangle \langle u_{c\mathbf{k}+e_\beta q} | u_{v\mathbf{k}} \rangle^*, \qquad (8)$$

where the indices $c$ and $v$ refer to conduction and valence band states respectively, and $u_{c\mathbf{k}}$ is the cell periodic part of the orbitals at the point $\mathbf{k}$ in the reciprocal space. The k-point weights $w_{\mathbf{k}}$ are defined according to the symmetry such that they sum to 1. The real part of the dielectric tensor is obtained from the imaginary part by the Kramers-Kronig transformation[45]

$$\varepsilon_{\alpha\beta}^{(1)}(\omega) = 1 + \frac{2}{\pi}\mathcal{P}\int_0^\infty \frac{\varepsilon_{\alpha\beta}^{(2)}(\omega')\omega'}{\omega'^2 - \omega^2 + i\eta}d\omega', \tag{9}$$

where $\mathcal{P}$ denotes taking the principal value. A small quantity $\eta$ is introduced by adding a factor into the perturbation of Hamiltonian $\delta\hat{H}(t) = e^{\eta t}\sum_{i=1}^N \delta V(\mathbf{r}_i, t)$ to guarantee an adiabatic switching from the unperturbed Hamiltonian ($t \to -\infty$, $\delta\hat{H}(t) \to 0$), in which $\delta V(\mathbf{r}_i, t)$ is the perturbation to the potential[44], and leads to an imaginary frequency shift after Fourier transform. This direct band summation method equation (8) is widely used for evaluating the macroscopic dielectric function of semiconductors and insulators[47,50,51]. However, it has some shortcomings: (i) only the direct transitions of electrons between valence and conduction bands are calculated, and the results include only the interband contribution, with the imaginary part of the DF having a direct connection with the electronic band structure; (ii) when deducing equation (8) from the linear response theory, the random phase approximation (RPA) is assumed, which ignored the electronic exchange-correlation effect (i.e., setting $f_{xc} = 0$ in the Dyson equation, see below)[49].

To obtain more accurate results, we also evaluate the DF using advanced method with the complete linear response theory in the second step. The main idea is to solve the response function $\chi$ within perturbation theory, which describes the induced changes in the density of electrons when a weak external electric field is applied to the material[49]. It is easier to calculate by first in the Kohn-Sham scheme, which is a system of independent electrons moving in a self-consistent effective potential. The relationship between the response function $\chi$ of an interacting electron system and $\chi^{KS}$ of the Kohn-Sham system is then given by the Dyson equation

$$\chi(\mathbf{r},\mathbf{r}',\omega) = \chi^{KS}(\mathbf{r},\mathbf{r}',\omega) + \int d^3r_1 d^3r_2 \chi^{KS}(\mathbf{r},\mathbf{r}_1,\omega)\left[\frac{e^2}{|\mathbf{r}_1-\mathbf{r}_2|} + f_{xc}(\mathbf{r}_1,\mathbf{r}_2,\omega)\right]\chi(\mathbf{r}_2,\mathbf{r}',\omega). \tag{10}$$

The second term in the right-hand side of equation (10) includes the Coulomb kernel, as well as the exchange-correlation kernel that can be approximated with different functional.

For weak metallic phases of cI16 and oC24, some bands are partially occupied. Therefore, we must consider the possible transitions from one state below the Fermi level to another state above the Fermi level within the same band. This intraband contribution

$$\varepsilon_{intra}^{(1)}(\omega) = -\frac{\bar{\omega}^2}{\omega^2}, \tag{11}$$

has to be added to the real part of the DF, where the plasmon frequency $\bar{\omega}$ has a form of[44]

$$\bar{\omega}_{\alpha\beta}^2 = \frac{4\pi e^2}{V}\sum_{n\mathbf{k}} 2f_{n\mathbf{k}}\frac{\partial^2 \epsilon_{n\mathbf{k}}}{\partial \mathbf{k}_\alpha \partial \mathbf{k}_\beta}. \tag{12}$$

The DF calculated in above methods corresponds to the microscopic one, from which the macroscopic DF that we are interested in can be obtained (which is also the quantity measured in experiment). The microscopic DF includes additional rapid oscillations of the total electric field on the scale of the primitive cell, whereas the macroscopic DF is homogenous on a coarse scale[52]. Generally, the macroscopic DF can be obtained by taking the long-wavelength limit and inverting the head ($\mathbf{G} = \mathbf{G}' = 0$) of the microscopic DF, which is a 3 × 3 tensor[49]

$$\varepsilon_{mac}(\omega, \hat{\mathbf{q}}) = (\lim_{q \to 0} \varepsilon_{00}^{-1}(\omega, \mathbf{q}))^{-1} . \tag{13}$$

The rapid oscillations on the microscopic scale caused by slowly varying external electric field and local crystalline structure may have an influence on the macroscopic response of the material to this field, and this is called local field effects (LFE), which represent the inhomogeneity of the electron gas. These are reflected in the off-diagonal elements of DF and have been automatically included in equation (13) in the second step calculations.

## B. Details of first-principles computation

The electronic structure and optical spectra calculations are performed using the *ab initio* total-energy and molecular-dynamics program VASP[53]. The projector-augmented wave (PAW) pseudopotential of Blöchl is employed, which combines the all-electron level accuracy with the computational efficiency of the pseudopotential approach, as implemented in VASP with the frozen-core approximation[54]. The longitudinal expression of the frequency-dependent microscopic dielectric function is utilized in the frame of PAW. The results are proved to be more accurate than the transversal expressions and are largely independent of the applied potential[49]. The generalized-gradient approximation (GGA) of Perdew-Burke-Ernzerhof (PBE) is used for the exchange-correlation functional of the density functional theory (DFT)[55]. All electrons ($1s^2 2s^1$) in lithium are treated as valence electrons. The electronic wave function is expanded using a plane wave basis up to an energy cutoff of 600 eV for ground-state calculations. An energy cutoff of 400 eV is used for optical spectra calculations, due to the high computational cost of the response method. Nevertheless, we have carefully checked and ensured the optical results have satisfactorily converged. Hellman-Feynman forces and stresses are systematically calculated and the ionic positions are fully optimized. All lattice vectors are also optimized subjected to the given hydrostatic pressure. In order to compute the interstitial charges, empty spheres are placed at the interstitial sites. The exact locations of their centers are approximately considered as the locations with the highest value of ELF in interstitial regions. We calculated the electronic DOS projected onto these spheres as the DOS of interstitial charges, where 95% is the *s* contribution. The Wigner radius of projection is set 0.8 Å, about half of the distance between the nearest ion and center of the interstitial charges, also the same as the projection radius on ions.

The first step of the DF calculation as mentioned above is directly based on the static results of the ground state DFT calculation, so it is labelled as 'interband part by DFT' hereinafter. The second step of the DF calculation is carried out by using the GW module as implemented in VASP, with an extra exchange-correlation kernel in the Dyson equation (10). In addition to the accurate macroscopic DF (labeled as 'GW with LFE' hereinafter), we also calculate the head of the microscopic DF in the GW module for comparison, which corresponds to the macroscopic DF in the independent particle approximation (IPA) (labeled as 'GW with IPA' hereinafter). These two results in the second step both include the intraband contributions and the exchange-correlation interaction in dielectric response. However, the former includes the LFEs, whereas the latter does not. The energy cutoff for the response function, which controls how many **G** vectors are included in the response function, is kept the same as the plane-wave cutoff energy. A total number of empty bands of 24 for cI16, 60 for oC40, and 36 for oC24 are employed in the optical calculations to converge the spectra up to ~10 eV. Because the transition probability highly depends on the accuracy of the integration over the k-space, an insufficient k-point sampling might introduce artificial errors into the optical spectra. Denser k-point samplings therefore have been utilized in the calculation of optical properties. A Monkhorst-Pack (MP) mesh[56] of $13 \times 13 \times 13$, $11 \times 11 \times 11$, and $13 \times 13 \times 13$ are used to sample the Brillouin zone for cI16, oC40, and oC24 in the band structure calculations and $17 \times 17 \times 17$, $11 \times 11 \times 11$, and $15 \times 15 \times 15$ in the optical spectra calculations, respectively. It gives a good convergence that has been carefully checked and confirmed. All symmetries have been switched off when dealing with the response function. A 0.1 eV Gaussian broadening of the optical spectra is applied. It should be pointed out that though more accurate optical properties could be obtained with the Bethe-Salpeter equation (BSE)[57,58], our results presented here are sufficiently accurate to give an quantitative optical spectra (see Supplementary Material), which is enough for our purpose to understand the evolution of the optical properties of dense lithium with increasing pressure.

## Acknowledgments

This work was supported by the National Natural Science Foundation of China under Grant 11672274, the NSAF under Grant U1730248, the Fund of National Key Laboratory of Shock Wave and Detonation Physics of China under Grant 6142A03010101, and the CAEP Research Project under Grant 2015B0101005. Part of the computation was performed using the supercomputer at the Center for Computational Materials Science of the Institute for Materials Research at Tohoku University, Japan.

## Author contributions

Zheng Yu performed the calculations, analyzed the data and wrote the manuscript. Hua Y. Geng provided the main idea, analyzed the data and wrote the manuscript. Y. Sun and Y. Chen contributed their intelligence in discussions.

## Competing financial interests

There is no conflict of interest.

## Data availability

All data generated or analyzed during this study are included in this published article (and its Supplementary Information files).

## Supplementary Information

It is provided online with the manuscript.

# Supplementary Material for "Optical properties of dense lithium in electride phases by first-principles calculations"

Zheng Yu[1], Hua Y. Geng[1], Y. Sun[1], and Y. Chen[2]

[1]*National Key Laboratory of Shock Wave and Detonation Physics, Institute of Fluid Physics, CAEP; P.O.Box 919-102 Mianyang, Sichuan P.R.China, 621900*

[2] *Fracture and Reliability Research Institute, School of Engineering, Tohoku University 6-6-01 Aramakiaoba, Aoba-ku, Sendai 980-8579, Japan*

## 1. Justification to the computational method

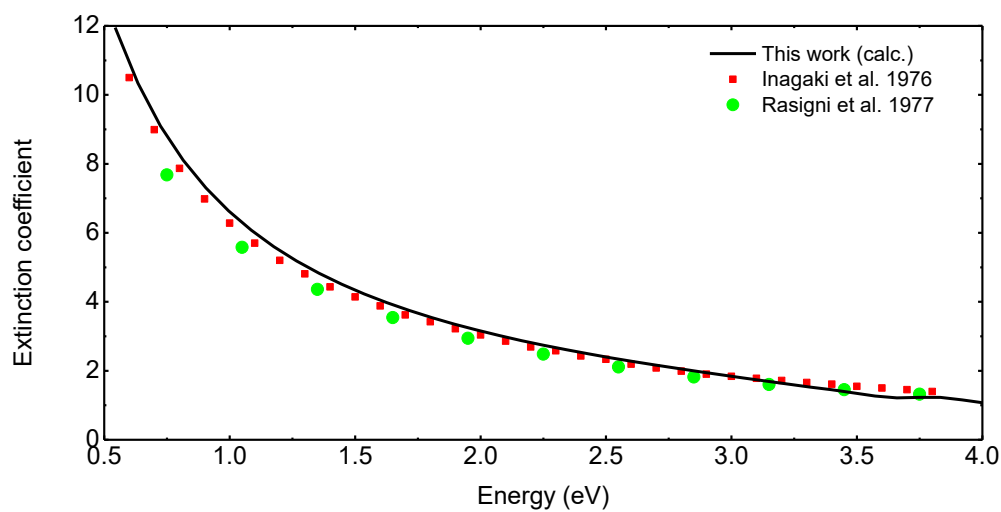

FIG. s12. Comparison of the calculated extinction coefficient of bcc lithium at 0 GPa in this work with the experimental data[1,2].

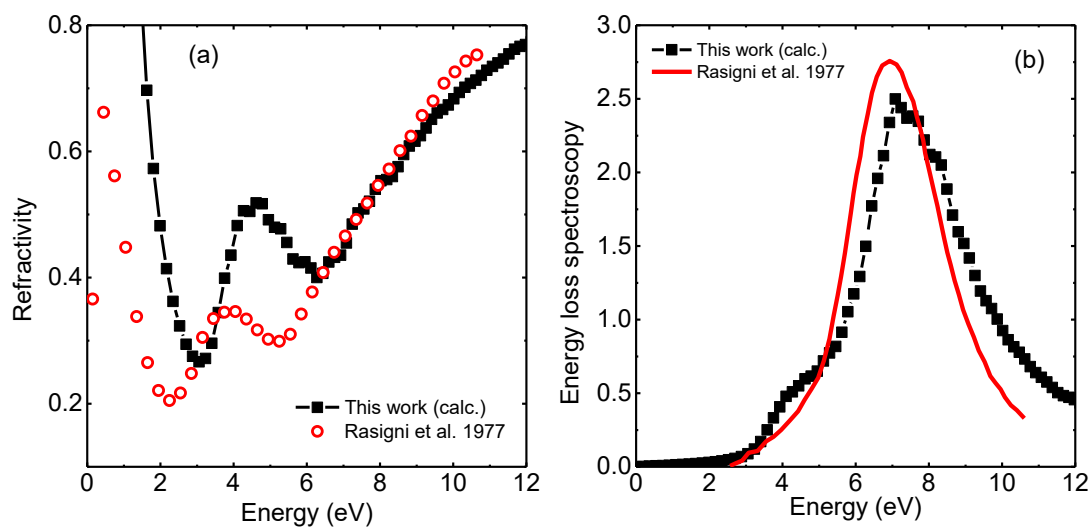

FIG. s13. Comparison of the calculated refractivity and the EELS of bcc lithium at 0 GPa in this work with the experimental results[2].

The validation and accuracy of the method for optical calculations used in this work has been carefully checked with the bcc structure of lithium at ambient conditions. The plasmon frequency of 7.0 eV for bcc lithium by our calculation is in good agreement with the experimental result of 6.73 eV[2]. The calculated energy loss function also matches the experimental spectrum very well, with a main peak of 7.0 eV against the measured 6.9 eV, as shown in Fig. s2(b). The calculated refractive index compares to the experimental data in Fig.2(a). The peaks in our results are slightly shifted to high frequency, but the overall feature agrees with the experimental spectrum. These lend us the confidence that our results are robust and quantitatively accurate.

Still note that the accuracy of optical calculations is in different levels with calculations of total energy. They should be considered as semi-quantitative, especially for the static electric resistivity. The exactly accurate calculation of static optical resistivity needs considerably dense k-points and huge supercells, so that numerous data in extremely low frequency regime could deduce resistivity for $\omega \to 0$. But this extremely high level of exactness is not required in this work. The data in Figure 11 of the main text is to show the metal-semiconductor-metal transition. We cannot determine whether the metallization of oC24 increases at higher pressure only from the subtle difference of resistivity between 115 GPa and 200 GPa.

In addition, VASP Li pseudopotential needs to be carefully checked. That's why we only consider the pressure range only up to 200 GPa. The latest PAW-PBE VASP pseudopotential of sv_GW with a core radius of 1.500 Bohr was used in this work. We checked its accuracy in a structure of $P4_132$ compared to the results obtained by all-electron full potential augment planewave (FP-LAPW)[3], as Table s1 shows. The pseudopotential of sv_GW shows reliable enough accuracy in calculation of volume of atoms at extreme pressures. The calculation of total energy is certainly affected by the pseudopotential. But the difference is evident only above 200 GPa (see Fig. 1(d) of Reference[3], where "VASP PAW PBE" means the old pseudopotential). The new sv_GW used in this work has better results. The electronic structure was also checked, in agreement with results by FP-LAPW method. In addition, our electronic structure calculation totally coincides with the results by Marques et al.[4]. Therefore, we believe the pseudopotential we used in this work is accurate enough for our optical research up to 200 GPa.

Table s1: Volumes of atoms calculated by different methods. The sv_GW pseudopotential is the one used in this work. Data of other methods are from Yao[3].

| Pressure (GPa) | 100 | 150 | 200 | 300 | 400 | 500 |
|---|---|---|---|---|---|---|
| VASP new sv_GW | 0.43% | 0.00% | -0.21% | -0.26% | 0.06% | 0.44% |
| QE TM PBE | 0.00% | 0.12% | 0.22% | 0.37% | 0.00% | 0.51% |
| VASP old PAW PBE | 1.15% | 1.01% | 1.03% | 2.32% | | 6.04% |
| ABINIT GTH PBE | 0.20% | 0.43% | 0.69% | 0.61% | | 1.05% |

## 2. Extra material for cI16

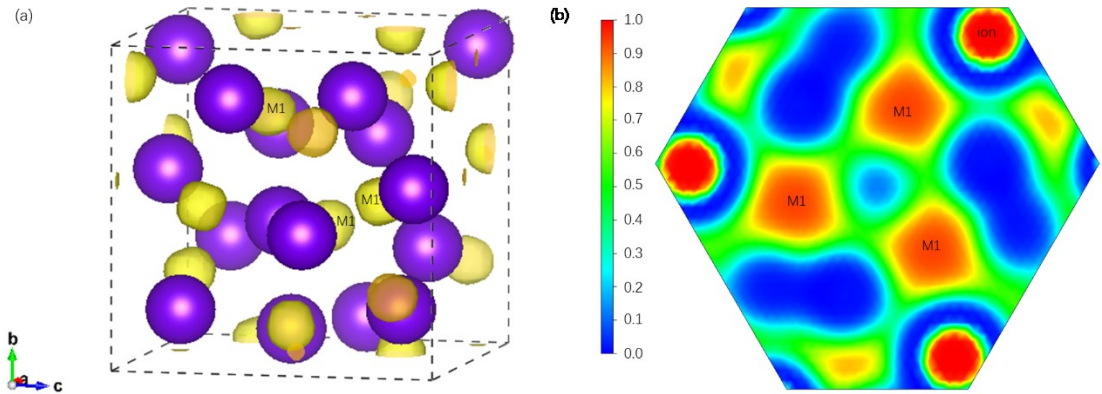

FIG. s14. (a) Structure of cI16 phase with an isosurface of ELF=0.9 at 70 GPa. Lithium ions are colored purple and the ELF isosurface is colored yellow. (b) A 2D display of the plane crossing the three interstitial electrons marked as M1 in (a). cI16 has only one type of ISQ.

The absence of the first peak in Reference[5] (see Fig. 2 of the main text) might be due to the premature cutoff of the extremely low-energy excitations (below 0.5 eV) around the Γ point, whose transition probability, however, is much larger than the high frequency ones. An artificial suppression up to 0.5 eV is evident in Reference[5]. The very close bands near the Fermi surfaces might have been treated as fully degenerate in Alonso's calculations, so their contribution to the interband part below 0.5 eV was ignored.

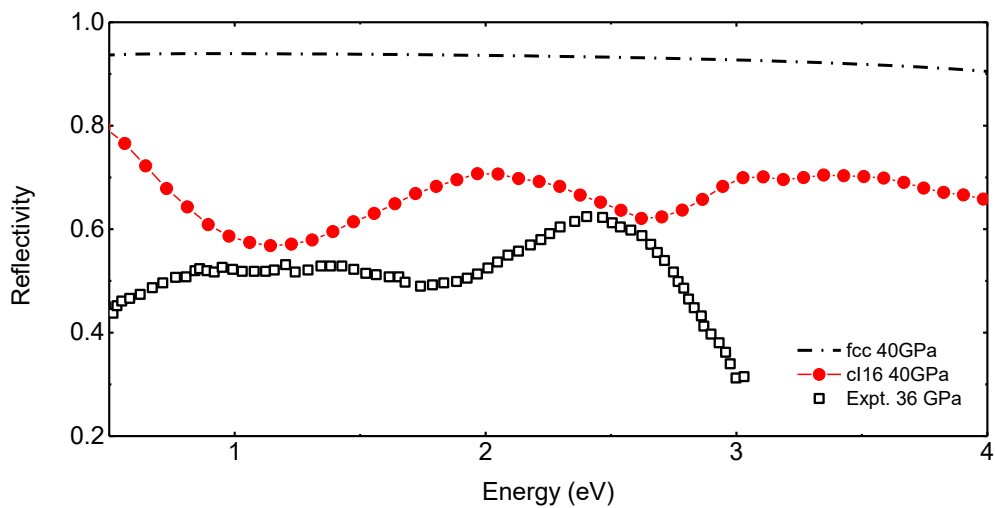

FIG. s15. Comparison of the calculated reflectivity of dense lithium in fcc and cI16 phase at about 40 GPa with experimental data[6]. The agreement between calculations and measurements is qualitative, and the possible reasons for the slight deviation have been discussed in the main text.

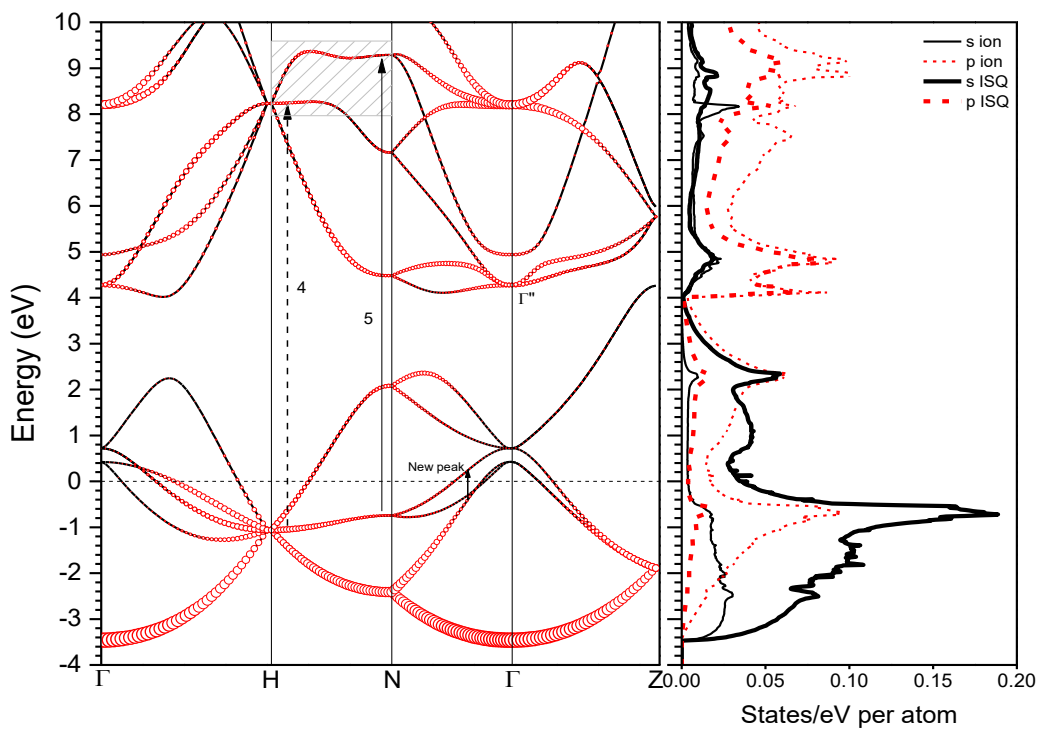

FIG. s16. Calculated electronic band structure and density of states of cI16 at 70 GPa. Other settings are the same as Fig. 1 in the main text. The band marked as $\Gamma'$ in Fig. 1 moves up to the position of $\Gamma''$ and forms a triply degenerate state by compression. The shadow rectangle region denotes the parallel bands which enhance the amplitude of the peaks 4 and 5 in the imaginary part of DF. The separating of the bands around $\Gamma$ point near the Fermi level leads to a new peak at 0.7 eV that is split out from the first main peak as shown in Fig. 2(a) of the main text.

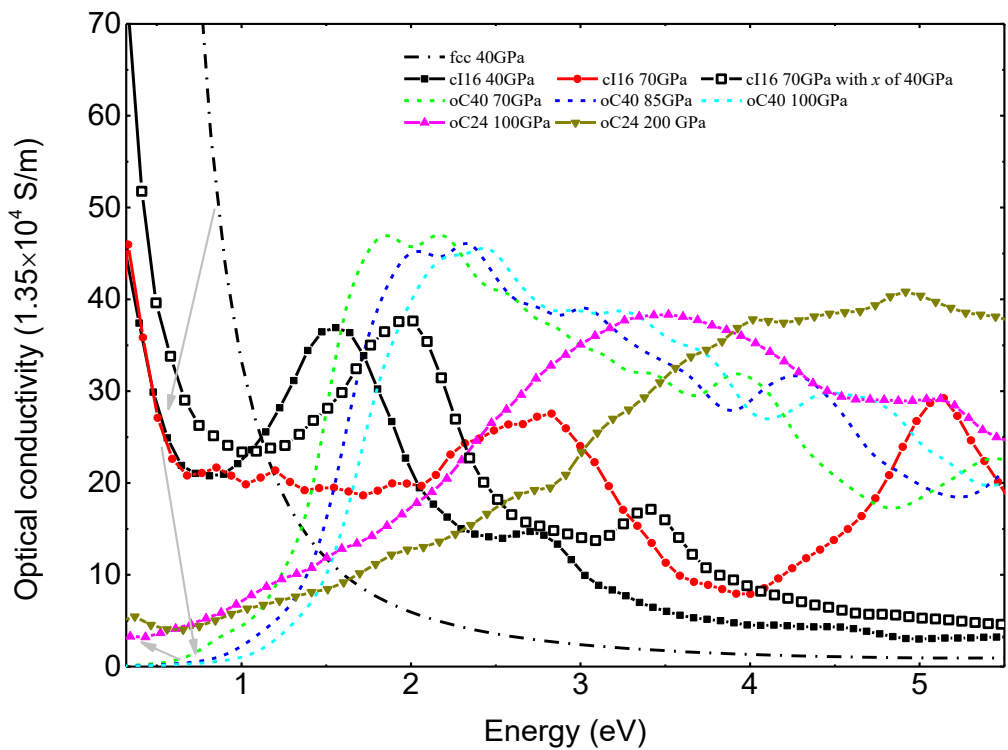

FIG. s17. Variation of the optical conductivity of dense lithium as a function of frequency in three electride phases under increasing pressure. The AC conductivity at low frequency illustrates the optical characteristics of the reentrant transition from metal to semi-metal, to insulator, and then back to a semi-metal phase.

## 3. Extra material for oC40

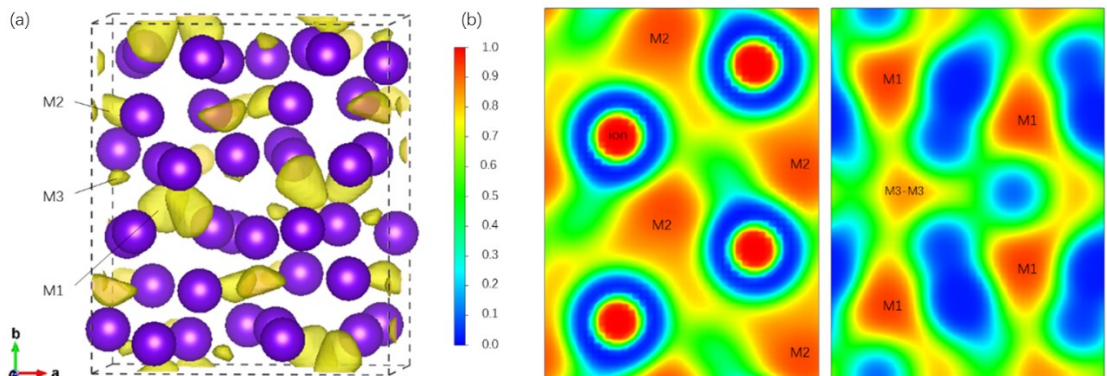

FIG. s18. (a) Structure of oC40 with an isosurface of ELF=0.88 at 70 GPa. Lithium ions are colored purple and the ELF isosurface is colored yellow. (b) 2D displays of the planes perpendicular to (010) crossing the interstitial electrons M1 and M2 marked in (a), respectively. The M3-M3 in the right panel of (b) denotes the chemical bond between adjacent M3s. oC40 has three types of ISQ.

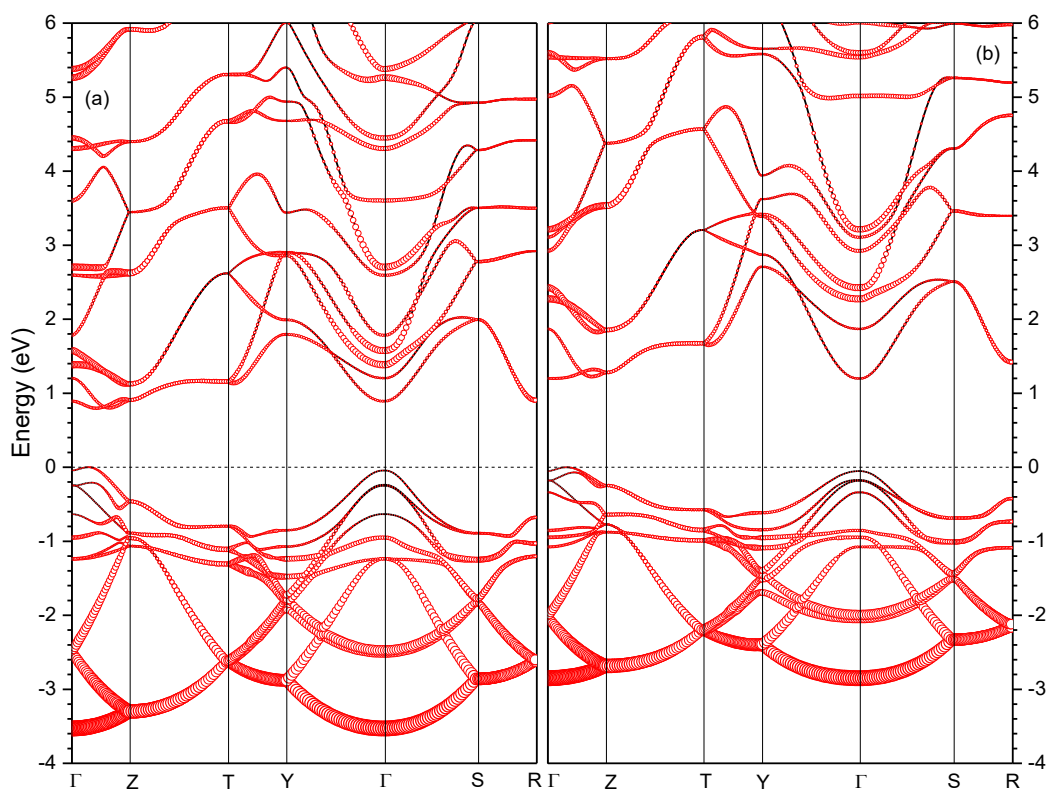

FIG. s19. Comparison of the band structure of oC40 at (a) 70 and (b) 100 GPa. The circle size is proportional to the band *s*-character. The valence bands are counterintuitively narrowed by pressure.

Insulating oC40 and semi-metallic cI16 have totally different symmetries, crystalline and electronic structures, and bonding. It's easy to understand that the electronic bands of oC40 evolve differently with pressure from that of cI16, especially not forming evident degenerate states at Γ point. In addition, as discussions in the manuscript indicate, the cI16 structure is distorted from the bcc structure, where only a few pairs of lithium ions in cI16 are separated by a distance while many others' locations remain unchanged. These unchanged ions also keep relatively unmoved when pressure rises from 40 to 70 GPa. However, the structure of oC40 is optimized more freely and completely at high pressure. All ions will move due to less symmetry constraints when pressure changes, only the distance of very special bonding atoms/quasi-atoms will keep unchanged[7]. Thus, the modification of electronic structure in oC40 is more intuitive, in which the conduction bands not far from the Fermi level are stretched due to different ratios of *2s* contribution. In cI16, due to unique structural distortion and the nesting structure, the electronic structure at Γ point alters abnormally, forming a triply-degenerate state at 70 GPa.

## 4. Extra material for oC24

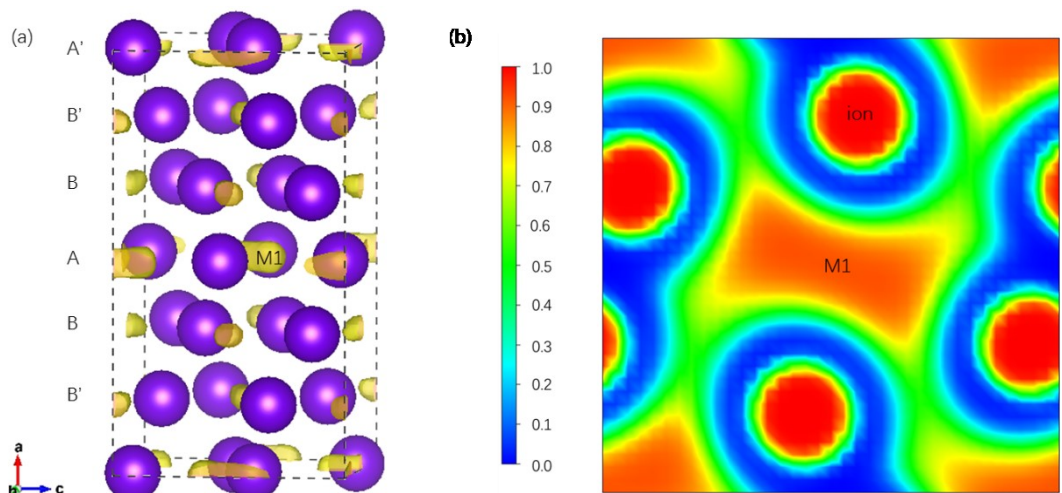

FIG. s20. (a) Structure of oC24 with an isosurface of ELF=0.88 at 100 GPa. Lithium ions are colored purple and the ELF isosurface is colored yellow. (b) A 2D display of the plane perpendicular to (100) crossing the interstitial electrons M1 as marked in (a). oC24 has two types of ISQ.

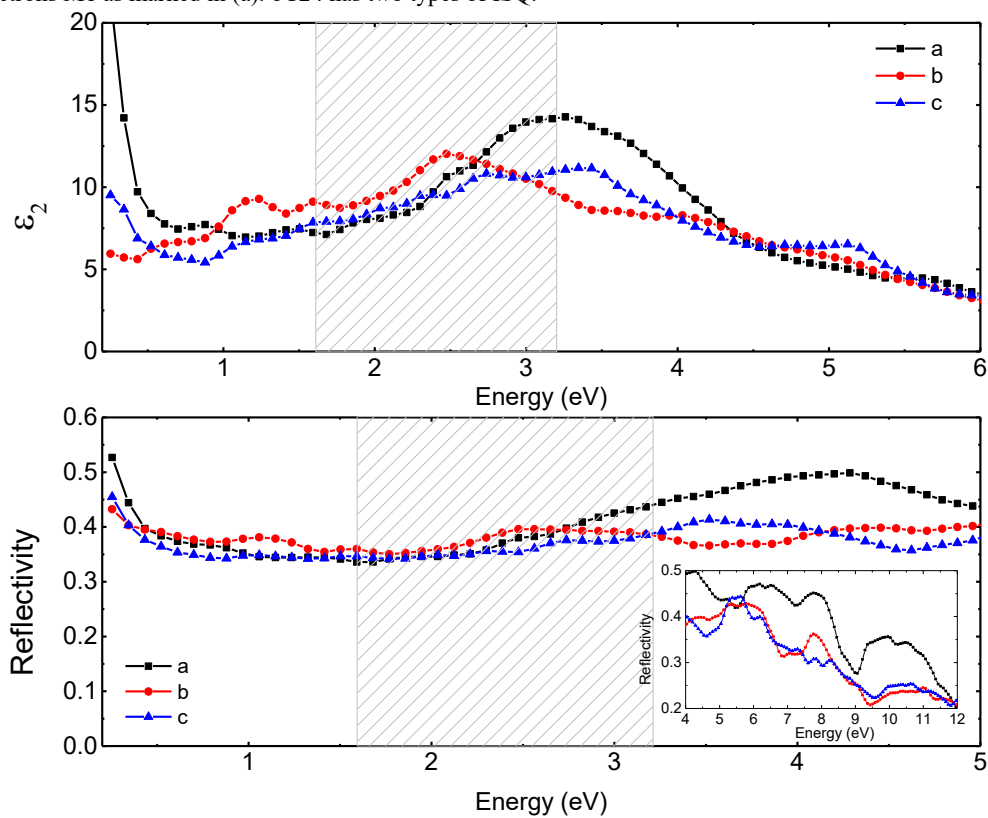

FIG. s21. The calculated anisotropy in the optical properties of oC24 at 100 GPa: (a) the imaginary part of DF; (b) the reflectivity. The anisotropy is weak in the visible light range (1.6-3.2 eV), but becomes noticeable in the ultraviolet regime.

oC24 phase inherits a layered structure from oC40, but has a higher symmetry, with a centrosymmetric distribution of atoms within each layer. Layers A′ and B′ are just another layer of A and B with relative shifts. oC24 has two non-equivalent interstitial electronic localization positions on layer A and B (ELF = 0.921 and 0.908 at 115 GPa), respectively. This structural feature accounts for the observed optical anisotropy.

The interband contribution to DF from electronic transitions with light polarization along *a* axis in the low energy region is much larger than along the other two directions, though this anisotropy disappears gradually under further compression. By the analysis in the main text, the first peak in the imaginary DF occurs within the cone around Γ point and this anisotropy represents an uneven feature of the cone structure. The actual momenta of the second peak differ in the three directions. It is at 3.2 eV along *a* axis, at 2.5 eV along *b* axis, and with two smaller peaks replacing the main one along *c* axis at 2.7 and 3.4 eV, respectively. However, the anisotropy in reflectivity is weak in infrared and visible light regime. But it becomes noticeable from 3 to 12 eV, as can be seen in Fig. s10. The reflectivity at 4 eV along *a* axis is about 0.1 higher than that along *b* and *c* axis, which becomes much more disparate at 7 eV and 10 eV.

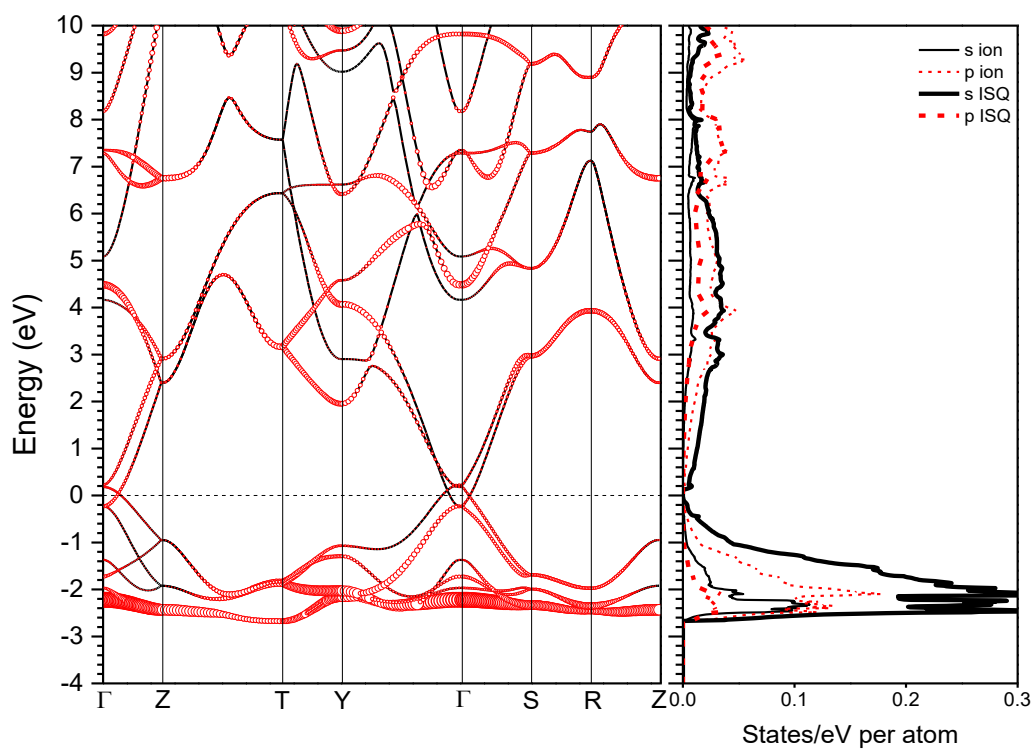

FIG. s22. Calculated electronic band structure and density of states of oC24 at 200 GPa. The main peaks of the *2s* and *2p* component overlap with each other exactly. The valence bands are consequently further narrowed, with striking localization of both *s* and *p* orbitals. But the cone structure is almost the same as oC24 at 100 GPa.

The local details of electronic structure of oC24 near the Fermi surface actually changes subtly with pressure, as shown in FIG 9 and FIG S11. Its cone feature has more area across the Fermi surface and the corresponding energy bands around Γ point also moves relatively. Although the metallization of oC24 could change subtly, it remains weakly metallic up to 200 GPa.

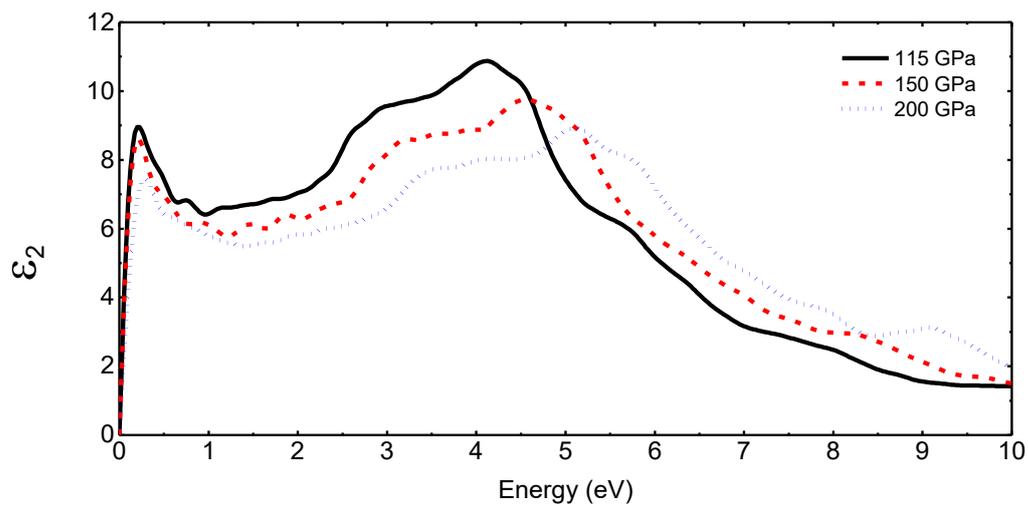

FIG. s23. Variation of the spherically averaged imaginary part of DF in oC24 at 115, 150, and 200 GPa, respectively.

# 5. Extra data of optical properties

Table s2. The reflectivity, refractive index, and absorption coefficient of high-pressure electride phases of lithium at pressures from 40 to 200 GPa at a frequency of 2, 3 and 4 eV, respectively. Note that oC24 is the most transparent one in the visible light range, even though it is metallic.

| 2 eV | pressure (GPa) | Reflectivity | Refractive index | Absorption coefficient |
|---|---|---|---|---|
| fcc | 40 | 0.94 | 0.34 | 8.8 |
| cI16 | 40 | 0.71 | 1.5 | 7.6 |
|  | 70 | 0.45 | 2.2 | 4.8 |
|  | 70 (*x*=0.048) | 0.60 | 2.6 | 7.2 |
| oC40 | 70 | 0.53 | 3.9 | 6.1 |
|  | 85 (along *a* axis) | 0.56 | 4.2 | 6.6 |
|  | 85 (along *b* axis) | 0.42 | 3.8 | 3.6 |
|  | 85 (along *c* axis) | 0.53 | 4.6 | 5.7 |
|  | 100 | 0.49 | 4.4 | 4.4 |
| oC24 | 100 | 0.35 | 3.4 | 2.6 |
|  | 200 | 0.31 | 3.1 | 2.0 |

| 3 eV | pressure (GPa) | Reflectivity | Refractive index | Absorption coefficient |
|---|---|---|---|---|
| fcc | 40 | 0.92 | 0.15 | 7.9 |
| cI16 | 40 | 0.70 | 0.70 | 7.6 |
|  | 70 | 0.54 | 1.6 | 8.0 |
|  | 70 (*x*=0.048) | 0.56 | 1.0 | 6.9 |
| oC40 | 70 | 0.50 | 2.2 | 8.1 |
|  | 85 (along *a* axis) | 0.51 | 2.4 | 8.5 |
|  | 85 (along *b* axis) | 0.39 | 2.6 | 6.1 |
|  | 85 (along *c* axis) | 0.53 | 2.4 | 8.9 |
|  | 100 | 0.46 | 2.6 | 7.5 |
| oC24 | 100 | 0.40 | 2.9 | 6.0 |
|  | 200 | 0.32 | 3.1 | 3.8 |

| 4 eV | pressure (GPa) | Reflectivity | Refractive index | Absorption coefficient |
|---|---|---|---|---|
| fcc | 40 | 0.90 | 0.10 | 6.8 |
| cI16 | 40 | 0.65 | 0.38 | 6.3 |
|  | 70 | 0.30 | 0.85 | 4.8 |
|  | 70 (*x*=0.048) | 0.61 | 0.59 | 7.4 |
| oC40 | 70 | 0.49 | 1.6 | 9.6 |
|  | 85 (along *a* axis) | 0.52 | 1.3 | 9.2 |
|  | 85 (along *b* axis) | 0.38 | 2.2 | 7.9 |
|  | 85 (along *c* axis) | 0.42 | 1.7 | 8.5 |
|  | 100 | 0.44 | 1.7 | 8.8 |
| oC24 | 100 | 0.43 | 2.0 | 8.9 |
|  | 200 | 0.36 | 2.6 | 7.2 |

## 6. ELF information on the three electride phases

Table s3: Bader charge and ELF analysis of dense lithium.

| Structure | fcc | cI16 | | oC40 | | oC24 | |
|---|---|---|---|---|---|---|---|
| Pressure (GPa) | 40 | 40 | 70 | 70 | 100 | 100 | 200 |
| Charge of each ion | 2.28 | 2.29 | 2.32 | 2.32 | 2.34 | 2.35 | 2.39 |
| ISQ charge occupation (from Bader analysis) | 23.8% | 23.6% | 22.8% | 22.7% | 22.0% | 21.8% | 20.4% |
| Maximum of ELF (interstitial) | 0.861 | 0.914 | 0.94 | 0.961 | 0.959 | 0.921 | 0.918 |
| Interstitial space occupation (ELF>0.9) | 0.00% | 0.49% | 2.53% | 2.46% | 2.90% | 0.65% | 0.47% |
| Interstitial space occupation (ELF>0.8) | 2.39% | 7.26% | 8.94% | 9.26% | 9.37% | 6.58% | 4.88% |

Table s3 is to aid us understand the electron distribution in the electride phases of lithium. The occupation of interstitial space with high ELF is calculated to represent the degree of the electron localization. The localization is greatly enhanced for cI16 from 40 to 70 GPa, as the occupation volume with ELF greater than 0.9 increases 5 times compared to other two phases. This striking change in the electron distribution results in a structural distortion of cI16, as discussed in the main text. The localization of oC24 is weakened at high pressure, which shows a weak metallic behavior.

## 7. Extra discussion

The *2p* contribution to the orbitals around ions certainly increases with pressure from 40 to 200 GPa and it plays a significant role in dense lithium. With the increase of *2p* contribution, the *pπ* bonding in cI16 becomes more saturated, allowing more valence electrons to localize at interstitial space. Thus, the structure of cI16 is gradually distorted with the increment of the internal coordinate *x*, creating more room in the interstitial region. Due to the nesting structure in cI16, there exists an energy gap between the highest not fully occupied valence band and the lowest conduction band. The gap becomes bigger with pressure and finally the structure stabilized by the *pπ* bonding is not stable anymore. The low-symmetric insulating oC40 has a lower enthalpy above 66 GPa due to the saturated multi-centered chemical bonding and the bonding between interstitial electrons of E1[7-9]. Under further compression the structure alters subtly, and the energy gap becomes larger, because the *2s* contribution at Γ-Z, Z-T, and T-Y makes the lowest conduction bands rises faster than the highest valence bands with the 2p contribution. As atomic structure becomes harder, the increasing of *2p* contribution becomes slower and it is not the main causes of these structural changes. The repulsion of interstitial electrons is more evident, making the energy reach the value for bond cleavage. Some interstitial electrons tend to be free. As a result, dense lithium transits into the reentrant weak metallic phase oC24. It can be confirmed by that the maximum of ELF in oC40 and oC24 decreases continuously with pressure.

As Figures 1, 6, and 9 in the main text show, the major peaks in the excitation spectra are related to those excitations from the highly occupied *2p* states not far below the Fermi surface to the corresponding *2s* states in conduction bands. As external pressure rises, though the *2p* contribution in valence bands increases, the *2s* states contributes more to the changes of excitation spectra, because they rise far faster than the *2p* states in band structure, resulting in the movement of the excitation spectra to higher frequencies. Except this, the changes of excitation spectra in cI16 is unique compared to oC40 and oC24, due to its structural distortion as we discussed in the manuscript, which is highly related to the *2p* contribution.